\documentclass[12pt]{article}
\usepackage{amsfonts}
\usepackage[cp1251]{inputenc}
\usepackage[english]{babel}
\usepackage{eufrak}
\begin{document}
 \sloppy
\title{Renormalization of Gauge Theories and the Hopf Algebra of
Diagrams }
\author{D.V. Prokhorenko \footnote{Institute of Spectroscopy, RAS 142190 Moskow Region, Troitsk}}
\maketitle
\begin{abstract}
 In 1999 A. Connes and D. Kreimer have discovered a Hopf algebra
 structure on the Feynman graphs of scalar field theory.
They have found that the renormalization can be interpreted as a
solving of some Riemann --- Hilbert problem. In this work the
generalization of
 their scheme to the case of nonabelian gauge theories is
 proposed. The action of the gauge group on the Hopf algebra of
 diagrams is defined and the proof that this action is in consistent
 with the Hopf algebra structure is given. The sketch of new proof
 of unitarity of \(S\) -matrix, based on the Hopf algebra approach
 is given.
\end{abstract}
\newpage
\section{Introduction}
 \indent The mathematical theory of renormalization (\(R\)-operation) was developed by
N.N. Bogoliubov and O.S. Parasiuk \cite{1}. K. Hepp has elaborated
on their proofs \cite{2}.

In 1999, A. Connes and D. Kreimer \cite{3,4} have discovered a
Hopf Algebra structure on the Feynman graphs in scalar field
theory with \(\varphi^3 \) interaction. The Hopf algebras play an
important role in the theory of quantum groups and other
noncommutative theories. (About noncommutative field theory and
its relation to p-adic analysis see \cite{5,6}.)

In the Connes --- Kreimer theory the Feynman amplitudes belongs to
the group of characters of the Hopf algebra of diagrams. Denote by
\(U\) a character corresponding to the set of nonrenormalized
amplitudes. Denote by \(R\) the character corresponding to the set
of renormalized amplitudes, and denote by \(C\) the character
corresponding to the counterterms. The following identity holds:
\begin{eqnarray}
R=C \star U.
\end{eqnarray}
Here, the star denotes the group operation in the group of
characters.

Denote by \(U(d)\) the dimensionally regularized Feynmann
amplitude (\(d\) is a parameter of dimensional regularization).
\(U(d)\) is holomorphic in a small neighborhood of the point
\(d=6\). We can consider \(U(d)\) as a data for the Riemann ---
Hilbert problem \cite{7} on the group of characters of Hopf
algebra of diagrams. A. Connes and D. Kreimer have proved that
this problem has an unique solution and the positive and negative
parts of The Birkhoff decomposition define renormalized amplitudes
and counterterms (if we use the minimal substraction scheme).
About future generalization of this scheme see \cite{8,9,10}.

In \cite{11} the generalization of this scheme to the case of
quantum electrodynamics is given. In gauge theories it is
necessary to prove that the renormalized Feynman amplitudes are
gauge invariant. In quantum electrodynamics the conditition of
gauge invariance is expressed in terms of the Ward identities and
in nonabelian gauge theories in terms of the Slavnov --- Tailor
identities.

Thus, an interesting problem is the problem of definition the
action of gauge group on the Hopf algebra of Feynman graphs such
that this action do not destroy the structure of Hopf algebra.

We solve this problem in the present paper.

Another the Hopf algebra description of renormalization theory of
nonabelian gauge fields was proposed in \cite{11'}.

The paper composed as follows. In section 2 we recall the basic
concept of Hopf algebras. In section 3 we define the algebra of
Feynman graphs (so-called Connes --- Kreimer algebra) and prove
that this algebra has an essential structure of Hopf algebras
(so-called generalized Connes --- Kreimer theorem).

In section 4 we recall the basic notion of gauge theories. In
section 5 we recall the continual integral method for quantization
gauge fields. In section 6 we derive the Slavnov --- Tailor
identities. Note that the usual Slavnov --- Tailor identities are
nonlinear, but our identities are linear. In section 7 we derive
the Slavnov --- Tailor identities for individual diagrams. In
section 9 we define the action of the gauge group on the Hopf
algebra of diagrams and prove our main results which state that
the action of gauge group do not destroy the Hopf Algebra
Structure. In section 10 we show how to apply our results to the
proof that physical observable quantities do not depend on the
special chose of gauge conditions.
\section{Hopf algebras}

 \textbf{Definition.} Coalgebra is a triple
${}\,(C,\Delta,\varepsilon)$, where $C$  is a linear space over
the field \(\mathbf{k}\); \(\Delta:C\rightarrow
C\),\({}\;\varepsilon: C \rightarrow \mathbf{k}\) are linear maps
satisfying the following axioms:\newline
 A)\begin{equation}
  (\Delta\otimes id)\circ \Delta= (id\otimes
 \Delta)\circ\Delta.
 \end{equation}
 B) The following map:
\begin{eqnarray}
 (id\otimes\varepsilon)\circ\Delta:C\rightarrow
 C\otimes\mathbf{k}\cong C, \\
 (\varepsilon\otimes id)\circ \Delta:C\rightarrow
\mathbf{k}\otimes C\cong C
\end{eqnarray}
are identical. The map \(\Delta\) is called a coproduct, and
\(\varepsilon\) is called a counity. The property A) is called a
coassociativity.

\textbf{Definition.} Coalgebra $(A,\Delta,\varepsilon)$ is a
bialgebra if \(A\) is an algebra and the comultiplication and
counit are homomorphism of algebras:
\begin{eqnarray}
\Delta(ab)=\Delta(a)\Delta(b),\;\Delta(\mathbf{1})=\mathbf{1}\otimes\mathbf{1},\\
\varepsilon(ab)=\varepsilon(a)\varepsilon(b),\;\varepsilon(\mathbf{1})=1.
\end{eqnarray}
\indent \textbf{Sweedler notation.} Let \((C,\Delta,\varepsilon)\)
be a coalgebra and let \(x\) be an element of \(C\).
 \(\Delta(x)\)
have the following form
\begin{eqnarray}
\Delta(x)=\sum \limits_{i} x'_i\otimes x''_i
\end{eqnarray}
for some \(x',x''\in C\). This sum can be formally rewritten as
follows
\begin{eqnarray}
\Delta(x)=\sum \limits_{(x)} x'\otimes x''.
\end{eqnarray}
This notations are called the Sweedler notation. In these terms
the coassociativity axiom can be rewritten as follows
\begin{eqnarray}
\sum \limits_{(x)}  (\sum \limits_{(x')} (x')'\otimes
(x')'')\otimes x''=\sum \limits_{(x)} x'\otimes(\sum
\limits_{(x'')} (x'')'\otimes (x'')'').
\end{eqnarray}
In Sweedler notation booth sides of these expressions can be
rewritten in the form
\begin{eqnarray} \sum \limits_{(x)} x'\otimes x'' \otimes x'''.
\end{eqnarray}

 \indent \textbf{Definition.} Let $(C,\Delta,\varepsilon)$ be
 a coalgebra, $A$ be an algebra. Let $f$, $g$ be linear maps
 $C\rightarrow A$; $f,g:C\rightarrow A$. By definition the
 convolution $f\star g$ of the maps $f$ and $g$ is the following map:
\begin{equation}
\mu\circ(f\otimes g)\circ\Delta:C\rightarrow A.
\end{equation}
Here \(\mu\) is an multiplication in $A$. $\mu:a\otimes b\mapsto
ab$.

 \textbf{Definition.} Let
$(A,\Delta,\varepsilon)$ be a bialgebra. The antipode map \(S\) in
this bialgebra is a linear map \(A\rightarrow A\) such that
\begin{equation}
S\star id=id \star S=\eta\circ\varepsilon.
\end{equation}
Here $\eta$ is a homomorphism $\mathbf{k}\rightarrow A$, $x\mapsto
\mathbf{1}x $ and $\mathbf{1}$ is a unit in $A$.

\indent \textbf{Definition.} Let $(A,\Delta,\varepsilon,S)$ be a
Hopf algebra over the field $\mathbf{k}$. Character \(\chi\) on
$A$ is an homomorphism $A\rightarrow\mathbf{k}$.

 Denote by $G$ the set of all characters. The product of two
 characters $\chi$ and $\rho$ is their convolution
 $\chi\star\rho$. One can check that $\chi\star\rho$ is an
 character. The convolution is associative. This fact follows from
 the coassociativity of \(\Delta\). There exists an identity \(\varepsilon\) in $G$.
 Indeed
\begin{eqnarray}
(\varepsilon\star\chi)(x)=\sum \limits_{(x)}
\varepsilon(x')\chi(x'')=\sum \limits_{(x)}
\chi(\varepsilon(x')x'')=\chi(\sum \limits_{(x)}
\varepsilon(x')x'')=\chi(x).
\end{eqnarray}
Thus we have proved that \(\varepsilon\) is a right identity.
Similarly one can prove that \(\varepsilon\) is a left identity.
For each \(\chi \in G\) there exists an inverse
$\chi^{-1}=\chi\circ S$. Indeed
\begin{eqnarray}
\chi\star(\chi\circ S)(x)=\sum \limits_{(x)}\chi(x')\chi(S(x''))=
\sum \limits_{(x)}\chi(x'S(x''))\nonumber\\
=\chi(\eta\circ\varepsilon(x))=\chi(\mathbf{1})\varepsilon(x)=\varepsilon(x).
\end{eqnarray}
 Similarly one can prove that $\chi\circ S$ is a left inverse of
 $\chi$. Therefore the following theorem holds.\newline
 \indent \textbf{Theorem 1.} \textsl{The set of all characters of a Hopf
 algebra is a group with respect the convolution as a
group operation.}

\indent \textbf{Example.} Let us consider the algebra \(H\) of all
polynomial functions on \(SL(2,\mathbb{C})\) with respect the
pointwise  multiplication. Then it is a Hopf algebra if we put:
\begin{eqnarray}
(\Delta F)(g_1,g_2)=F(g_1g_2),\nonumber\\
\varepsilon(F)=F(e),\nonumber\\
(S(F))(g)=F(g^{-1}).
\end{eqnarray}
Here \(g_1\:,g_2\:,g\) are elements of \(SL(2,\mathbb{C})\), F is
a polynomial function on \(SL(2,\mathbb{C})\), \(e\) is the
identity in \(SL(2,\mathbb{C})\).

The group of characters \(G\) of \(H\) is isomorphic to
\(SL(2,\mathbb{C})\). This isomorphism to each element \(g\) of
\(SL(2,\mathbb{C})\) assigns a character \(\chi_g\), defined as
\begin{eqnarray}
\chi_g(F)=F(g),\; F \in H.
\end{eqnarray}
\indent \textbf{Definition.} Let
$C_1=(A_1,\Delta_1,\varepsilon_1)$ and
$C_1=(A_2,\Delta_2,\varepsilon_2)$ be coalgebras. A homomorphism
from $C_1$ to $C_2$ is a linear map $f:A_1\rightarrow A_2$ such
that
\begin{eqnarray}
\Delta_2\circ f=(f\otimes f)\circ \Delta_1 ,\\
\varepsilon_2\circ f=\varepsilon_1.
\end{eqnarray}
\indent \textbf{Definition.} Let
$H_1=(A_1,\Delta_1,\varepsilon_1,S_1)$ and
$H_2=(A_2,\Delta_2,\varepsilon_2,S_2)$ be Hopf algebras. The
homomorphism $f:A_1\rightarrow A_2$ is a Hopf algebra homomorphism
$f:H_1\rightarrow H_2$ if \(f\) is a coalgebra homomorphism
$f:C_i\rightarrow C_i$, where $C_i=(A_i,\Delta_i,\varepsilon_i)$
$(i=1,2)$, and
\begin{eqnarray}
S_2\circ f=f\circ S_1.
\end{eqnarray}
 \indent As usual in the case of Hopf algebra we can define the composition
 of homomorphisms and define the monomorphisms, epimorphisms etc. \newline
\indent \textbf{Definition.} Let $H=(A,\Delta,\varepsilon,S)$ be a
Hopf algebra. A derivation \(\delta\) of the Hopf algebra $H$ is a
derivation of $A$ such that
\begin{eqnarray}
\Delta\circ\delta=(id\otimes\delta+\delta\otimes id)\circ \Delta,\\
\varepsilon\circ\delta=0,\\
S\circ\delta=\delta\circ S.
\end{eqnarray}
\indent \textbf{Remark.} We can think about the derivatives as
about the infinitesimal automorphism.
\section{Feynman Diagrams} Let us define the Feynman diagrams.
Suppose that the theory describes \(N\) fields
\(\Phi_a^{\alpha}\), where \(a=1,...,N\) is an index numerating
different fields, \(\alpha\) ia an index, numerating different
components of fields. (This index may be spinor, vector, group
etc.) \(\alpha=1,...,\alpha_a\). For each field corresponding to
the index \(a\) we assign its index space $\mathbb{Z}_a:
 =\mathbb{C}^{\alpha_a}$ ($\mathbb{Z}_a:
 =\mathbb{R}^{\alpha_a}$).\newline
 \indent \textbf{Definition.} A Feynman graph is a triple
 ${\Phi}=(V,\{R_a\}_{a=1}^{N},f)$, where
 \(V\) is a finite set, called a set of vertices, and
 $\forall a =1,...,N$ \(R_a\) is a finite set, called a set of lines for the
  particles of type \(a\).
  Put by definition $R=\bigcup\limits_{a=1}^N R_a$. \(f\) is a map
 $f:R\rightarrow V\times V\cup V\times \{+,-\}$ \newline
 \indent \textbf{Definition.} Let \(r \in R\) be a line \(r\in f^{-1}(V\times V)\)
 or equivalently \(f(r)=(v_1,v_2)\) for some vertecies \(v_1\) and
 \(v_2\). We say that the line \(r\) comes into the vertex \(v_1\)
 and
 comes from the vertex \(v_2\). We say also that the vertecies \(v_1\) and \(v_2\)
 are connected by the line \(r\).

 Let \(r\) be a line such that  \(f(r)=(v,+)\). We say that the line \(r\)
 is an external line coming from the vertex \(v\). We also say
 that the line \(r\) comes from the Feynman graph \(G\).

Let \(r\) be a line such that  \(f(r)=(v,-)\). We say that the
line \(r\)
 is an external line coming into the vertex \(v\). We also say
 that the line \(r\) comes into the Feynman graph \(G\).

\indent \textbf{Definition.} The Feynman graph \(\Phi\) is called
connected if for two any vertecies  \(v,v'\) there exists a
sequence of vertecies \(v=v_0,v_1,...,v_n=v'\) such that
 \(\forall\, i=0,...,n-1\) the vertecies
 \(v_i\) and \(v_{i+1}\) are connected by some line. \newline
 \indent \textbf{Definition.} A Feynman graph \(\Phi\) is called
 one particle irreducible if it is connected and can not be
 disconnected by removing a single line.

 \indent Let \({\Phi}\) be a Feynman graph. Let
 \(v\) be a vertex of \({\Phi}\). We let \(R^{\rightarrow v}\) be
 a set of all lines coming into the vertex \(v\), and \(R^{\leftarrow v}\)
 be a set of all lines coming from the vertex
  \(v\). Let:
 \begin{eqnarray}
 \mathbb{Z}_v=\{\bigotimes \limits_{r \in R^{\leftarrow v}} \mathbb{Z}_{a_r}\}\bigotimes
 \{\bigotimes \limits_{r \in R^{\rightarrow v}}
 \mathbb{Z}_{a_r}^\star\}.
 \end{eqnarray}
 Here \(a_r\) is a type of particle, corresponding to the line
 \(r\).
 \(V^\star\) is a dual of the space \(V\).\newline
 \indent \textbf{Definition.} The space \(\mathbb{Z}_v\)
 is called an index space of the vertex \(v\).\newline
\indent \textbf{Definition.} Let \({\Phi}\) be a Feynman graph,
and \(v\) be a vertex of \({\Phi}\). The space \({S}_v\) is a
space of all linear combinations of the function of the form

 \begin{eqnarray}
\delta(\sum \limits_{r\rightarrow v}p_r-p) f(p_r).
\end{eqnarray}
Here \(f(p_r)\) is an arbitrary polynomial of variables \(\{p_r|r
\in R^{\leftarrow v}\cup R^{\rightarrow v}\}\) whose range is
\(\mathbb{Z}_v\).

 \textbf{Definition.} The Feynman
diagram is a pair $\Gamma=(\Phi,\varphi)$, where
${\Phi}=(V,\{R_a\}_{a=1}^{N},f)$ is a Feynman graph and $\varphi$,
is a map which assigns to each vertex \(v \in V\) an element
\(\varphi(v)\) of \(S_v\).
\newline
 \indent
 We will write below
 ${\Phi}_\Gamma$, \(\varphi_\Gamma\), to point out that the
 Feynman graph \(\Phi\) and the function \(\varphi\) corresponds
 to the diagram \(\Gamma\).

\textbf{Definition.} Let \({\Gamma}=(\Phi,\varphi)\) be a diagram
\begin{eqnarray}
{\Phi}=(V,\{R_a\}_{a=1}^{N},f) \nonumber
\end{eqnarray}
and \(I\) be a set of all its external lines. Let \(L_\Gamma\) be
a set of all maps \(I\rightarrow\mathbb{R}^4\), \(i\mapsto p(i)\).
\(L_\Gamma\) is called a space of external particle momenta.

Let \({\Gamma}=({\Phi},\varphi)\) be a Feynman diagram. Let
\(R^{\rightarrow {\Gamma}}\) be a set of all external lines of
\(\Phi\) coming into \(\Phi\). Let \(R^{\leftarrow {\Gamma}}\) be
a set of external lines of \(\Phi\) coming from \(\Phi\). Let
 \begin{eqnarray}
 \mathbb{Z}_{{\Gamma}}=\{\bigotimes \limits_{r \in R^{\leftarrow {\Gamma}}} \mathbb{Z}_{a_r}\}\bigotimes
 \{\bigotimes \limits_{r \in R^{\rightarrow {\Gamma}}}
 \mathbb{Z}_{a_r}^\star\}.
 \end{eqnarray}
 Here \(a_r\) is a type of particle corresponding to \(r\),
and \(V^\star\) is a dual space of the space \(V\).\newline
 \indent \textbf{Definition.} The space \(\mathbb{Z}_{{\Gamma}}\)
 is called an index space of the diagram \({\Gamma}\).\newline
 \indent \textbf{Definition.} \(S_\Gamma\) is a space of all
 linear combination of the functions of the form

  \begin{eqnarray}
  \delta(\sum \limits_{{r \in R^{\rightarrow\Gamma} \cup R^{\leftarrow\Gamma}}}p_r-p) f(p_r),
  \end{eqnarray}
  Here \(f(p)\) is a polynomial map
from \(L_{{\Gamma}}\)
 to \(\mathbb{Z}_{{\Gamma}}\).\newline
 \indent \textbf{Definition.} Let \({S_\Gamma}'\) be a algebraic
 dual of the space \({S_\Gamma}'\).
\({S_\Gamma}'\) is called a space of external structure of
\(\Gamma\).\newline
 \indent \textbf{Definition.}
  Let \(\mathcal{H}\) be a commutative unital algebra
  generated by the pairs \((\Gamma,\sigma)\) ($\Gamma$ is one
  particle irreducible diagram, $\sigma \in {S_\Gamma}'$) with the
  following
  relations
  \begin{eqnarray}
  (\Gamma,
  \lambda\sigma'+\mu\sigma'')=\lambda(\Gamma,\sigma')+\mu(\Gamma,\sigma''),\nonumber\\
  (\lambda\Gamma'+\mu\Gamma'',\sigma)=\lambda(\Gamma',\sigma')+\mu(\Gamma'',\sigma). \nonumber
\end{eqnarray}
Here \(\Gamma'\), \(\Gamma''\) and \(\lambda\Gamma'+\mu\Gamma''\)
are the diagrams such that
\begin{eqnarray}
\Phi_{\Gamma'}=\Phi_{\Gamma''}=\Phi_{\lambda\Gamma'+\mu\Gamma''}
\end{eqnarray}
and there exists  a vertex \(v_0\) of \(\Phi_{\Gamma'}\) such that
\begin{eqnarray}
\varphi_{\Gamma'}(v)=\varphi_{\Gamma''}(v)=\varphi_{\lambda\Gamma'+\mu\Gamma''}(v)\;\rm
if\; \mit
v\neq v_0  \nonumber\\
\varphi_{\lambda\Gamma'+\mu\Gamma''}(v_0)=\lambda
\varphi_{\Gamma'}(v_0)+\mu\varphi_{\Gamma''}(v_0).\nonumber\\
\end{eqnarray}
\(\mathcal{H}\) is called an algebra of Feynman diagrams. \indent

Let us give some notation necessary to give a definition of
coproduct on the algebra of Feynman diagrams.

Let \(B_\Gamma=\{l^\alpha_\Gamma\}\), \(\alpha \in
\mathrm{A}_\Gamma\) be an arbitrary Hamele basis of a space
\(S_\Gamma^{\Omega}\). Denote by
\(B_\Gamma^{'}=\{l^{\alpha'}_{\Gamma}\}\) the dual basis of
\(B_\Gamma=\{l^\alpha_\Gamma\}\).

 \textbf{Definition.} Let
 \(\Gamma=(\Phi,\varphi )\) be a one particle irreducible Feynman diagram, where
 \(\Phi=(V,\{R_a \}_{a=1}^{N},f)\). Let \(V'\) be a subset of
  \(V\). Let \( \tilde{R'}_a\) be a subset of \(R_a\) for each \(a=1...N\),
such that \(\forall r \in \tilde{R'}_a\) there exists vertecies
\(v_1\) and \(v_2\) from \(V'\) connected by \(r\).

Let \(\tilde{R}''_a\) be a subset of \( (R_a \setminus
\tilde{R'_a})\times \{+,-\} \),
\(\tilde{R''_a}:={\tilde{R}}^{''+}_a \cup {\tilde{R}}^{''-}_a\).
Here \(\tilde{R}^{''+}_a\)
 is a set of all pairs \((r,+)\) such that \(r \in R_a \setminus
 \tilde{R'_a}\) and \(r\) comes from \(V'\), \(\tilde{R}^{''-}_a\)
 is a set of all pairs \((r,-)\) such that \(r \in R_a \setminus
 \tilde{R'_a}\) and \(r\) comes into \(V'\). Put by definition
 \(R_a'=\tilde{R}'_a\cup \tilde{R}''_a\).

 \indent Let
\(\Phi_\gamma=(V',\{R'_a\}_{a=1}^{N},f')\) be a Feynman Graph,
where \(V'\), \(R'_a\) are just defined and
 \(f'(r):=f(r)\), if \(r \in \tilde{R}^{'}_a\),
 \(f'((r,+))=(v,+)\) if \((r,+) \in \tilde{R}^{''+}_a\) and
 \(f(r)=(v',v)\)
 or \(f(r)=(v,+)\);
 \(f'((r,-))=(v,-)\), if \((r,-) \in \tilde{R}^{''-}_a\) and
 \(f(r)=(v,v')\) or \(f(r)=(v,-)\).
 Let \(\gamma:=(\Phi_\gamma,
 \varphi_\gamma)\), where \(\varphi_\gamma\) is a restriction of
 \(\varphi_\Gamma\) to \(V'\). If \(\Phi_\gamma\) is one particle
 irreducible diagram \(\gamma\) is called an one particle
 irreducible subdiagram of
 \(\Gamma\).

  \textbf{Definition.} Let \(\gamma=\{\gamma_i|i=1,...,n\}\) be a set of one particle
  irreducible subdiagrams of \(\Gamma\) such that \(V_{\gamma_i}\cap
  V_{\gamma_j}\neq\emptyset\) \(\forall i\neq j\). We say that
  \(\gamma\) is a subdiagram of \(\Gamma\). \(\forall i=1,...,n\) \(\gamma_i\) is called a connected
  component of \(\gamma\). Let \(M=\{1,...,n\}\).
  The elements of \(M\) numerate the connected components of
  \(M\). Let \(\alpha\) be a map which to each element of \(M\)
  assigns the element \(\alpha(i)\) of \(\mathrm{A}_{\gamma_i}\).
  \(\alpha\) is called a multi index. Let \(\gamma' \) be a
  subdiagram of \(\Gamma=(\Phi,\varphi)\) and \(\alpha\) be a
  multi index. We assign to the pair \((\gamma',\alpha)\) an
  element \( \gamma_{\alpha} \):=\( \prod \limits_{i \in M }(
  \gamma_{i},l_{\gamma_i}^{\alpha(i)'})\) of \(\mathcal{H}\).

  The quotient diagram \(\Gamma/\gamma_{\alpha}\) as a graph  is
  obtained by replacing each of the connected component
  \(\gamma_i\) of \(\gamma\) by the corresponding vertex \(v_i\).
  For each \(i \in M\) we can identify \(S_{\gamma_i}\) with
  \(S_{v_i}\). We put by definition
  \(\varphi_{\Gamma/\gamma_\alpha}(v)=\varphi(v)\) if \(v\neq v_i\)
  \(\forall i \in M\) and

  \(\varphi_{\Gamma/{\gamma_\alpha}}(v_i)=l^{\alpha(i)}_{\gamma_i}\).

  \textbf{Definition.} Comultiplication \(\Delta\) is a
  homomorphism \(\mathcal{H}\rightarrow \mathcal{H}\otimes \mathcal{H}\),
  defined on generators as follows:
  \begin{equation}
  \Delta ((\Gamma,\sigma))=(\Gamma,\sigma)\otimes \mathbf{1}+
  \mathbf{1} \otimes (\Gamma,\sigma) + \sum \limits_{ \emptyset \subset
  \gamma_{\alpha} \subset \Gamma
  } {\gamma_{\alpha}} \otimes
  ( {\Gamma}/{\gamma_{\alpha}},\sigma),
  \label{Delta}
  \end{equation}
  (see. \cite{3,4}.)

   \textbf{Remark.} In the previous formula \(\subset\) means the strong inclusion. The sum is over all
   nonempty subdiagrams \( \gamma \subset
  \Gamma \) and multiindecies \(\alpha\).

  \textbf{Theorem 2.} \textsl{The homomorphism \(\Delta\) is well defined and do not depend of
  a special chose of a basis \(B_{\Gamma}\) of \(S_{\Gamma}\)}.\newline
  \indent \textbf{Proof.} It is evidence.\newline
  \indent \textbf{Theorem 2. (The generalized Connes ---  Kreimer theorem.)} \textsl{ Homomorphism \(\Delta\)
  is coassociative. Moreover we can find a counit \(\varepsilon\)
  and an antipode \(S\) such that \((\mathcal{H},\Delta,\varepsilon,S)\)
  is a Hopf algebra.}

  \textbf{Proof.}
  Let \(\Gamma\) be a Feynman diagram and
  \(\gamma_\alpha,\gamma_\beta\) are subdiagrams of \(\Gamma\)
  such that \(\gamma_\alpha\subset\gamma_\beta\). We can define a
  quotient diagram \(\gamma_\beta/\gamma_\alpha\) by the evident
  way.

  Let us show that \(\Delta\) is coassociative. We have:
  \begin{equation}
  \Delta((\Gamma,\sigma))=(\Gamma,\sigma)\otimes\mathbf{1}+\mathbf{1}\otimes(\Gamma,\sigma)
  + \sum \limits_{ \emptyset \subset
  \gamma_{\alpha} \subset \Gamma
  } {\gamma_{\alpha}} \otimes
  ( {\Gamma}/{\gamma_{\alpha}},\sigma),
  \end{equation}
  \begin{eqnarray}
  (\Delta\otimes
  id)\circ\Delta((\Gamma,\sigma))=(\Gamma,\sigma)\otimes\mathbf{1}\otimes\mathbf{1}+
  \mathbf{1}\otimes(\Gamma,\sigma)\otimes\mathbf{1}+\mathbf{1}\otimes\mathbf{1}\otimes(\Gamma,\sigma)
  \nonumber\\
  +\sum \limits_{ \emptyset \subset
  \gamma_{\alpha} \subset \Gamma
  } {\gamma_{\alpha}} \otimes
  (
  {\Gamma}/{\gamma_{\alpha}},\sigma)\otimes\mathbf{1}\nonumber\\
+\sum \limits_{ \emptyset \subset
  \gamma_{\alpha} \subset \Gamma
  } \Delta({\gamma_{\alpha}}) \otimes
  ( {\Gamma}/{\gamma_{\alpha}},\sigma).
  \end{eqnarray}
\begin{eqnarray}
(\Delta\otimes
  id)\circ\Delta((\Gamma,\sigma))=(\Gamma,\sigma)\otimes\mathbf{1}\otimes\mathbf{1}+
  \mathbf{1}\otimes(\Gamma,\sigma)\otimes\mathbf{1}+\mathbf{1}\otimes\mathbf{1}\otimes(\Gamma,\sigma)
  \nonumber\\
   +\sum \limits_{ \emptyset \subset
  \gamma_{\alpha} \subset \Gamma
  } {\gamma_{\alpha}} \otimes
  (
  {\Gamma}/{\gamma_{\alpha}},\sigma)\otimes\mathbf{1}+
  \sum \limits_{ \emptyset \subset
  \gamma_{\alpha} \subset \Gamma
  } {\gamma_{\alpha}} \otimes \mathbf{1}\otimes
  (
  {\Gamma}/{\gamma_{\alpha}},\sigma)
  \nonumber\\
  +\mathbf{1}\otimes\sum \limits_{ \emptyset \subset
  \gamma_{\alpha} \subset \Gamma
  } {\gamma_{\alpha}} \otimes
  (
  {\Gamma}/{\gamma_{\alpha}},\sigma)+
\sum \limits_{ \emptyset \subset \gamma_{\beta}\subset
  \gamma_{\alpha} \subset \Gamma
  } {\gamma_{\beta}}\otimes{\gamma_{\alpha}}/ {\gamma_{\beta}}\otimes
  (
  {\Gamma}/{\gamma_{\alpha}},\sigma).\label{1}
  \end{eqnarray}
  From other hand:
  \begin{eqnarray}
  (id\otimes\Delta)\circ\Delta((\Gamma,\sigma))\nonumber\\
  =(id\otimes\Delta)\{
  (\Gamma,\sigma)\otimes\mathbf{1}+\mathbf{1}\otimes(\Gamma,\sigma)
  + \sum \limits_{ \emptyset \subset
  \gamma_{\alpha} \subset \Gamma
  } {\gamma_{\alpha}} \otimes
  ( {\Gamma}/{\gamma_{\alpha}},\sigma)\}\nonumber\\
  =(\Gamma,\sigma)\otimes\mathbf{1}\otimes\mathbf{1}+
  \mathbf{1}\otimes(\Gamma,\sigma)\otimes\mathbf{1}+\mathbf{1}\otimes\mathbf{1}\otimes(\Gamma,\sigma)
  \nonumber\\
  +\mathbf{1}\otimes\sum \limits_{ \emptyset \subset
  \gamma_{\alpha} \subset \Gamma
  } {\gamma_{\alpha}} \otimes
  (
  {\Gamma}/{\gamma_{\alpha}},\sigma)\nonumber\\
  +\sum \limits_{ \emptyset \subset
  \gamma_{\alpha} \subset \Gamma
  } {\gamma_{\alpha}} \otimes
  (
  {\Gamma}/{\gamma_{\alpha}},\sigma)\otimes\mathbf{1}
  +\sum \limits_{ \emptyset \subset
  \gamma_{\alpha} \subset \Gamma
  } {\gamma_{\alpha}} \otimes \mathbf{1}\otimes
  (
  {\Gamma}/{\gamma_{\alpha}},\sigma)
  \nonumber\\
  +\sum \limits_{\emptyset \subset
  \gamma_{\alpha} \subset \Gamma;\emptyset \subset
  \gamma_{\beta} \subset \Gamma/\gamma_\alpha}
  \gamma_\alpha\otimes\gamma_\beta\otimes((\Gamma/\gamma_\alpha)/\gamma_\beta,\sigma)
  \label{2}
  \end{eqnarray}
  To conclude the prove of the theorem it is enough to prove
  the coincidence of the last terms of (\ref{1}) and (\ref{2}). In other words it is enough to
  prove the following equality
\begin{eqnarray}
\sum \limits_{ \emptyset \subset \gamma_{\beta}\subset
  \gamma_{\alpha} \subset \Gamma
  } {\gamma_{\beta}}\otimes{\gamma_{\alpha}}/ {\gamma_{\beta}}\otimes
  (
  {\Gamma}/{\gamma_{\alpha}},\sigma)\nonumber\\
=\sum \limits_{\emptyset \subset
  \gamma_{\gamma} \subset \Gamma;\emptyset \subset
  \gamma_{\delta} \subset \Gamma/\gamma_\gamma}
  \gamma_\gamma\otimes\gamma_\delta\otimes((\Gamma/\gamma_\gamma)/\gamma_\delta,\sigma)
  \label{3}
  \end{eqnarray}
  To each term of left hand side of (\ref{3})
\begin{equation}
\gamma_{\beta}\otimes{\gamma_{\alpha}}/ {\gamma_{\beta}}\otimes
  (
  {\Gamma}/{\gamma_{\alpha}},\sigma)
  \end{equation}
assign the following term of the right hand side of (\ref{3})
\begin{eqnarray}
\gamma_\gamma\otimes\gamma_\delta\otimes((\Gamma/\gamma_\gamma)/\gamma_\delta,\sigma)
  \label{3},
\end{eqnarray}
where \(\gamma=\beta\),
\(\gamma_\delta=\gamma_\alpha/\gamma_\beta\). It is evidence that
this map is a bijection and
\(\Gamma/{\gamma_\alpha}=(\Gamma/{\gamma_\gamma})/\gamma_\delta\).
So the equality (\ref{3}) holds. The coassociativity of \(\Delta\)
is proved.\newline
 \indent It is easy to see that the homomorphism \(\varepsilon:\mathcal{H}\rightarrow\mathbb{C}\)
 defined by
 \(\varepsilon((\Gamma,\sigma))=0,\:\rm if \mit\;\Gamma\neq\emptyset\),
 \(\varepsilon(\mathbf{1})=1\) is a counit in \(\mathcal{H}\)

 Let \(\tilde{\mathcal{H}}\) be a linear subspace of
 \(\mathcal{H}\) spanned by the elements \(\mathbf{1}\) and \(\{(\Gamma,\sigma)\}\). Let
 us define the linear function
 \(S:\tilde{\mathcal{H}}\rightarrow\mathcal{H}\) by using the following
 reccurent relations
\begin{equation}
S((\Gamma,\sigma))=-(\Gamma,\sigma)-\sum
\limits_{\emptyset\subset\gamma_\alpha\subset\Gamma}(\gamma_\alpha)S((\Gamma/{\gamma_\alpha},\sigma)).
\label{Ant}
\end{equation}
The order of the diagrams in the right hand side less than \(n\)
if the order of \(\Gamma\) is equal to \(n\).

Now let us extend \(S\) to a map \(S:\mathcal{H}\rightarrow
\mathcal{H}\) by the following rule
\begin{equation}
S((\Gamma_1,\sigma_1)...(\Gamma_n,\sigma_n))=S((\Gamma_1,\sigma_1))...S((\Gamma_n,\sigma_n)).
\end{equation}
One can prove that just defined map
\(S:\mathcal{H}\rightarrow\mathcal{H}\) is an antipode in
\(\mathcal{H}\). The Theorem is proved.

 \textbf{Definition.}
Let \(\Gamma\) be an one particle irreducible Feynman diagram. Let
\(C_\Gamma\) be a space of all \(\mathbb{Z}_\Gamma\)-valued
distributions on \(L_{{\Gamma}}\) which are finite linear
combinations of the distributions of the form
\begin{eqnarray}
  \delta(\sum \limits_{r \in R^{\rightarrow\Gamma}\cup R^{\leftarrow\Gamma}}p_r-p)
  f(p_r).
\end{eqnarray}
Here \(f(p_r)\) is an arbitrary \(\mathbb{Z}_\Gamma\)-valued
smooth function with compact support on \(L_{{\Gamma}}\). Let
\(C'_\Gamma\) be an algebraic dual of \(C_\Gamma\). Let \(M\) be a
linear space spanned by the pairs \((\Gamma,\sigma)\), \(\sigma
\in C'_\Gamma\) with relation expressing the linearity of
\((\Gamma,\sigma)\) by \(\Gamma\) and \(\sigma\).  One can prove
that
  \(M\) is a comodule over \(\mathcal{H}\) if one define the
  comultiplication on \(\mathcal{H}\) by the formula
 (\ref{Delta}).

\section{The Yang --- Mills action}
\indent Let \(G\) be a compact semisimple Lie Group,
\(\mathfrak{g}\) be its Lie algebra and \(\hat{}\) be its adjoint
representation. It is possible to find a basis of \(\mathfrak{g}\)
(a set of generators) \(\{T^a\}\) such that
\begin{equation}
\langle T^a T^b\rangle\equiv {\rm tr \mit}
\hat{T}^a\hat{T}^b=-2\delta^{ab}.
\end{equation}
\indent \textbf{Defimition.} Gauge field is a
\(\mathfrak{g}\)-valued one-form on \(\mathbb{R}^4\):
\begin{equation}
A=\sum \limits_{\mu=1}^4 \sum \limits_a A^a_\mu dx^\mu T^a.
\end{equation}
\indent \textbf{The covariant derivative.} Let \(\Gamma\) be a
representation of \(G\) by complex \(n\times n\) matricies acting
in  \(V=\mathbb{C}^n\).

 \textbf{Definition.} Let \(R\) be a trivial bundle over
 \(\mathbb{R}^4\) with the fibre \(V\).

 Let \(A_\mu\) be a gauge field. The covariant derivative \(\nabla_\mu\) is a map
 \begin{eqnarray}
 \nabla_\mu:\Gamma(R)\rightarrow \Gamma(R)
 \end{eqnarray}
of the form
\begin{equation}
\nabla_\mu\psi=\partial_\mu\psi-g\Gamma(A_\mu)\psi,\;\psi \in
\Gamma(R).
\end{equation}
Here \(\Gamma(R)\) is a space of global sections of \(R\).

\textbf{Curvature.} Let \(A\) be a gauge field. Its curvature are
defined as
\begin{eqnarray}
{F}_{\mu\nu}=\partial_\nu A_\mu-\partial_\mu A_\nu+g[A_\mu,A_\nu].
\end{eqnarray}
One can easily check that
\begin{eqnarray}
[\nabla_\mu,\nabla_\nu]=g\Gamma {(F_{\mu\nu})}.
\end{eqnarray}

\textbf{Gauge transformation.} Let \(\omega(x)\) be a smooth map
from \(\mathbb{R}^4\) to \(G\). Gauge transformation is an
automorphism of \(R\) defined as
\begin{eqnarray}
 \psi(x)\rightarrow \psi'(x)=\Gamma(\omega(x))\psi(x).
\end{eqnarray}

 Under the gauge transformation \(\omega(x)\) the field
\(A\) transforms as follows
\begin{eqnarray}
A\rightarrow A'_\mu=\omega A_\mu
\omega^{-1}+(\partial_\mu\omega)\omega^{-1}.
\end{eqnarray}
This rule follows from the formula
\begin{eqnarray}
\nabla'_\mu
\Gamma(\omega(x))\psi(x)\}=\Gamma(\omega(x))\{\nabla_\mu\psi(x)\},\: \mbox{where}\\
\nabla'_\mu=\partial_\mu-\Gamma(A'_\mu).
\end{eqnarray}

 The curvature \(F\) under gauge transformations transforms as
 follows
\begin{eqnarray}
F\rightarrow F'=\omega F\omega^{-1}.
\end{eqnarray}

\indent \textbf{The Yang --- Mills action.} Let \(\Gamma^a\) be an
element of \(\mathfrak{g}^{\mathbb{C}}\) (complexification of
\(\mathfrak{g}\)) such that \(T^a=i\Gamma^a\), where
\(i=\sqrt{-1}\). We have

\begin{eqnarray}
{\rm tr \mit} (\hat{\Gamma}^a,\hat{\Gamma}^b)=2\delta^{ab}.
\end{eqnarray}
By definition
\begin{eqnarray}
[\Gamma^a,\Gamma^b]=if^{abc}\Gamma^c.
\end{eqnarray}

One can rewrite the curvature $F=F^aT^a$ as follows
\begin{eqnarray}
F^a_{\mu\nu}=\partial_\nu {A}^a_\mu-\partial_\mu
{A}^a_\nu-gf^{abc}{A}^b_\mu {A}^c_\nu.
\end{eqnarray}

The pure Yang --- Mills action by definition has the form
\begin{eqnarray}
S_{YM}[A]=-\frac{1}{8} \int \langle F_{\mu\nu},F_{\mu\nu}\rangle
d^4x=\frac{1}{4}F^a_{\mu\nu}F^a_{\mu\nu}d^4x.
\end{eqnarray}

The action for fermions has the form
\begin{eqnarray}
S_F=\int\bar{\psi}(i\gamma_\mu\nabla_\mu+m)\psi d^4x.
\end{eqnarray}
Here \(\gamma_\mu\) are the Euclidean Dirac matricies. The action
for the fermion interacting with the gauge field has the form
\begin{eqnarray}
S=S_{YM}+S_F.
\end{eqnarray}
The action \(S\) is  an ivariant under the gauge transformation if
the fermions under the gauge transformation transform as follows
\begin{eqnarray}
 \psi\rightarrow\psi'=\omega\psi, \nonumber\\
 \bar{\psi}\rightarrow\bar{\psi}'=\bar{\psi}\omega^{-1}.
\end{eqnarray}
\section{Quantization of the Yang --- Mills theory}
Let us recall the quantization procedure of the Yang --- Mills
theory by using the continual integral method.

Let $G[A,\bar{\psi},\psi]$ be a gauge invariant functional, i.e.
$G[A,\bar{\psi},\psi]$ satisfies
\begin{eqnarray}
G[{}^\omega
A,{}^\omega\bar{\psi},{}^\omega\psi]=G[A,\bar{\psi},\psi],
\end{eqnarray}
where
\begin{eqnarray}
{}^\omega A:= \omega A \omega^{-1}+(\partial_\mu
\omega)\omega^{-1},\nonumber\\
{}^\omega\psi:=\omega\psi\nonumber,\\
{}^\omega\bar{\psi}:=\psi\omega^{-1}.
\end{eqnarray}
The expectation value of the functional \(G[A,\bar{\psi},\psi]\)
by definition can be expressed through the continual integral as
follows
\begin{eqnarray}
\langle G[A,\bar{\psi},\psi]\rangle=\mathcal{N}^{-1} \int
DAD\bar{\psi}D\psi G[A,\bar{\psi},\psi] e^{-S[A]}.
\end{eqnarray}

Here \(\mathcal{N}\) is a constant such that
\(\langle1\rangle=1\).

This integral contains the integration over the gauge group. Our
aim is to include the volume of the gauge group into the
\(\mathcal{N}\).

Let $\chi[A](x)$ be a \(\mathfrak{g}\)-valued function on
\(\mathbb{R}^4\) depending of \(A\)
(${\chi[A](x)}={i\chi^a[A](x)\Gamma^a}$). \(\chi[A](x)\) are
called gauge functions. By definition the gauge surface is a set
of all field configurations \((A,\bar{\psi},\psi)\) such that
\(\chi[A](x)=0\;\forall x \in \mathbb{R}^4\). We suppose that the
gauge conditions are nondegenerate i.e.
\begin{eqnarray}
\rm det \mit\left \|\frac{\delta\chi^a[{}^\omega
A](x)}{\delta\omega^b(y)}\right \|\neq 0
\end{eqnarray}
if \(A\) belongs to the gauge surface. Let $\Delta[A]$ be a gauge
invariant functional such that
\begin{eqnarray}
\Delta [A]\int D \omega \delta(\chi[{}^\omega A])=1.
\end{eqnarray}
We have
\begin{eqnarray}
\Delta [A]=\rm det \mit\left \|\frac{\delta\chi[{}^\omega
A]}{\delta\omega}\right \|
\end{eqnarray}
if the field configuration \((A,\bar{\psi},\psi)\) lies on the
gauge surface. We have
\begin{eqnarray}
\langle
G[A_\mu,\bar{\psi},\psi]\rangle\nonumber\\
=\mathcal{N}^{-1}\int DAD\bar{\psi}D\psi \int D\omega
\delta(\chi[{}^\omega A])\Delta[A]
e^{-S[A,\bar{\psi},\psi]}G[A,\bar{\psi},\psi].
\end{eqnarray}
The functional \(G\), the action \(S\), the measure \(DA D\psi
D\bar{\psi}\) and the functional \(\Delta[A]\) are gauge
invariant, therefore after the changing variables
\begin{eqnarray}
\bar{\psi},\psi,A\rightarrow
{}^{\omega^{-1}}\bar{\psi},{}^{\omega^{-1}}\psi,{}^{\omega^{-1}}A
\end{eqnarray}
we can rewrite the last formula as follows
\begin{eqnarray}
\langle G[A\bar{\psi},\psi]\rangle\nonumber\\
=\mathcal{N}^{-1}\int D\omega \int DAD\bar{\psi}D\psi \delta(\chi[
A])\rm det \mit\left \|\frac{\delta\chi[{}^\omega
A]}{\delta\omega}\right \|
e^{-S[A,\overline{\psi},\psi]}G[A,\overline{\psi},\psi].
\label{FP}
\end{eqnarray}
Now we can include the integral \(\int D\omega\) into the
multiplier \(\mathcal{N}^{-1}\).

\textbf{The Faddeev --- Popov ghosts.} By definition the Faddeev
--- Popov ghosts are two \(\mathfrak{g}\)-valued Grassman fields
\(c^a(x)\) and \(\bar{c}^a(x)\). We have
\begin{eqnarray}
{\rm det \mit}\left \|\frac{\delta\chi[{}^\omega
A]}{\delta\omega}\right \|= \int D\bar{c}Dc
e^{\int\bar{c}^a(y)\frac{\delta\chi^a[{}^\omega
A](x)}{\delta\omega^b(y)} c^b(x)dxdy}.
\end{eqnarray}
Now let us use a new gauge conditions
$\chi^{a'}[A](x)=\chi^a[A](x)-f^a(x)=0$ in (\ref{FP}) instead of
$\chi^a[A](x)=0$, where $f^a$ is an arbitrary
\(\mathfrak{g}\)-valued function and integrate both sides of
(\ref{FP}) over \(f^{a}\) with a weigh $e^{-\frac{1}{2}\int
f^a(x)f^a(x)dx}$. In result we have
\begin{eqnarray}
\langle G[A,\bar{\psi},\psi]\rangle=\mathcal{N}^{-1} \int
DAD\bar{\psi}D\psi D\bar{c}Dc \,G[A,\bar{\psi},\psi]
e^{-\{S_{YM}+S_F+S_{FP}+S_{GF}\}},
\end{eqnarray}
where
\begin{eqnarray}
S_{FP}=-\int\bar{c}^a(y)\frac{\delta\chi^a[{}^\omega
A](x)}{\delta\omega^b(y)} c^b(x)dxdy
\end{eqnarray}
and
\begin{eqnarray}
S_{GF}=\frac{1}{2}\int(\chi^a[A](x))^2.
\end{eqnarray}
If we use the Lorentz gauge condition  \(\partial_\mu A_\mu=0\)
then we have
\begin{eqnarray}
S_{FP}=\int\partial_\mu\bar{c}^a(y)\nabla_\mu c^a=-\frac{1}{2}\int
\langle \partial_\mu\bar{c},\nabla_\mu c \rangle.
\end{eqnarray}
By definition, under the gauge transformation the ghosts
transforms as follows
\begin{eqnarray}
\bar{c}\mapsto\bar{c},\nonumber\\
c\mapsto \omega c \omega^{-1}.
\end{eqnarray}
\section{The Slavnov --- Taylor identities} Here we derive the
Slavnov --- Taylor identities. Note that our Slavnov --- Taylor
identities are linear but the usual Slavnov --- Taylor identities
are nonlinear.

 \textbf{The Green functions}. Let us use the Lorenz gauge conditions.
 The Green functions are defined as
\begin{eqnarray}
\langle A(x_1)...A(x_n)
\bar{\psi}(y_1)...,\bar{\psi}(y_m)\psi(z_1)...\psi(z_k)
\rangle\nonumber\\
=\int DAD\bar{\psi}D\psi D\bar{c} Dc\,e^{-S}A(x_1)...A(x_n)
\bar{\psi}(y_1)...,\bar{\psi}(y_m)\psi(z_1)...\psi(z_k).
\end{eqnarray}
 The generating functional for the Green functions are defined as
\begin{eqnarray}
Z[J,\bar{\eta},\eta]=\int DAD\bar{\psi}D\psi D\bar{c}
Dc\,e^{-S+\langle J,A\rangle+\langle
\bar{\eta},\psi\rangle+\langle \bar{\psi},{\eta}\rangle},
\end{eqnarray}
where
\begin{eqnarray}
 \langle J,A\rangle:= \int J^a_\mu A^a_\mu d^4x, \nonumber\\
\langle \bar{\eta},\psi \rangle:=\int \bar{\eta}\psi d^4x,\nonumber\\
\langle \bar{\psi},{\eta}\rangle:=\int \bar{\psi}{\eta} d^4x.
\end{eqnarray}
Now we can calculate the Green functions as the functional
derivatives of \(Z[J,\bar{\eta},\eta]\).

The generating functional for the connected Green functions are
defined as
\begin{eqnarray}
F[J,\bar{\eta},\eta]={\rm ln \mit}Z[J,\bar{\eta},\eta].
\end{eqnarray}
At last, the generating functional for the one particle
irreducible Green functions are defined by using the Legendre
transformation
\begin{eqnarray}
-\Gamma[A,\bar{\psi},\psi]=\langle J,A\rangle+\langle
\bar{\eta},\psi\rangle+\langle
\bar{\psi},{\eta}\rangle-F[J,\bar{\eta},\eta],
\end{eqnarray}
where \(J,\bar{\eta},\eta\) satisfy the conditions
\begin{eqnarray}
A=\frac{\delta}{\delta J} F[J,\bar{\eta},\eta],\nonumber\\
\bar{\psi}=-\frac{\delta}{\delta {\eta}}F[J,\bar{\eta},\eta],\nonumber\\
\psi=\frac{\delta}{\delta {\bar{\eta}}}F[J,\bar{\eta},\eta].
\label{72}
\end{eqnarray}
\indent \textbf{The Slavnov --- Taylor identities} Let
\(\omega=1+\alpha\) be an infinitezimal gauge transformation. Let
us compute the following expression:
\begin{equation}
\delta_\omega\Gamma[{}^\omega A,{}^\omega \bar{\psi},{}^\omega
\psi]:=\Gamma[{}^\omega A,{}^\omega \bar{\psi},{}^\omega
\psi]-\Gamma[ A, \bar{\psi}, \psi].
\end{equation}
We have
\begin{eqnarray}
-\delta_\omega\Gamma[{}^\omega A,{}^\omega \bar{\psi},{}^\omega
\psi]=\langle J,\delta_\omega A\rangle+\langle
\delta_\omega\bar{\psi},\eta\rangle+\langle\bar{\eta},\delta_\omega{\psi}
\rangle\nonumber\\
 +\langle \delta_\omega J,
A\rangle+\langle
\bar{\psi},\delta_\omega\eta\rangle+\langle\delta_\omega\bar{\eta},{\psi}
\rangle-\delta_\omega F[J].
\end{eqnarray}
The conditions (\ref{72}) implies that
\begin{eqnarray}
-\delta_\omega\Gamma[{}^\omega A,{}^\omega \bar{\psi},{}^\omega
\psi]=\langle J,\delta_\omega
A\rangle+\langle\bar\delta_\omega{\psi},\eta\rangle+\langle\bar{\eta},\delta_\omega{\psi}
\rangle.
\end{eqnarray}
From other hand we have:
\begin{eqnarray}
1=\frac{Z[J]}{Z[J]}=\frac{1}{Z[J]}\int DAD\bar{\psi}D\psi
D\bar{c}Dce^{-S+ \langle J,A\rangle+\langle
\bar{\eta},\psi\rangle+\langle \bar{\psi},{\eta}\rangle}.
\end{eqnarray}
It follows from the gauge invariance of the measure that
\begin{eqnarray}
0=\frac{1}{Z[J]}\int DAD\bar{\psi}D\psi D\bar{c}Dc e^{\{-S+
\langle J,A\rangle+\langle \bar{\eta},\psi \rangle+\langle
\bar{\psi},{\eta}\rangle \}}\nonumber\\
 \{\langle J,\delta_\omega
A\rangle+\langle \bar{\eta},\delta_\omega\psi\rangle+\langle
\delta_\omega\bar{\psi},{\eta}\rangle-\delta_\omega S\}.
\label{Var}
\end{eqnarray}
Let us introduce the following notation
\begin{eqnarray}
S_\omega[A,\bar{\psi},\psi]:= S[{}^\omega
A,{}^\omega\bar{\psi},{}^\omega\psi].
\end{eqnarray}
Let
$Z_\omega[J,\bar{\eta},\eta],\;F_\omega[J,\bar{\eta},\eta],\;\Gamma_\omega[A,\bar{\psi},\psi]$
 be generating functionals corresponding to the action
 \(S_\omega\). It follows from (\ref{Var}) that
\begin{eqnarray}
\langle\delta_\omega {}^\omega  A, J\rangle+\langle
\bar{\eta},\delta_\omega\psi\rangle+\langle
\delta_\omega\bar{\psi},{\eta}\rangle+\delta_\omega
F_\omega[J,\bar{\eta},\eta]=0,
\end{eqnarray}
but
\begin{eqnarray}
\delta_\omega\Gamma_\omega[A,\bar{\psi},\psi]=+\delta_\omega
F_\omega [J,\bar{\eta},\eta].
\end{eqnarray}
Therefore
\begin{eqnarray}
\delta_\omega\Gamma [{}^\omega A,
{}^\omega\bar{\psi},{}^\omega\psi]=\delta_\omega \Gamma_\omega
[A,\bar{\psi},\psi].
\end{eqnarray}

These equations we call the Slavnov --- Taylor identities.\newline
\section{The Feynman rules for the Yang --- Mills theory} We use the Lorenz gauge condition.
The action has the form
\begin{eqnarray}
S=\int\{\frac{1}{4} {\rm tr \mit}
F_{\mu\nu}^aF_{\mu\nu}^a+\frac{1}{2}(\partial_\mu
A^a)^2+\partial_\mu\bar{c}^a(\partial_\mu
c-g[A_\mu,c])^a\nonumber\\
+\bar{\psi}(i\gamma_\mu \nabla_\mu+m)\psi\}d^4x.
\end{eqnarray}
The quadratic part of the action has the form
\begin{eqnarray}
S_2=\int\{\frac{1}{2}(\partial_\nu A_\mu^a)^2+\partial_\mu
\bar{c}^a\partial_\mu {c}^a+\bar{\psi}(i\gamma_\mu
\partial_\mu+m)\psi\}d^4x.
\end{eqnarray}
Let us write the terms describing the interaction. \newline

 \textbf{The four-gluon interaction} is described by the
 following vertex
\begin{eqnarray}
V_{4A}=-\frac{g^2}{4} \int  [A_\mu,A_\nu]^a[A_\mu,A_\nu]^a d^4x\nonumber\\
=-\frac{g^2}{4} \int  f^{abe}f^{cde}A_\mu^a A_\nu^b A_\mu^c
A_\nu^d
 d^4x.
\end{eqnarray}

\textbf{The three gluon interaction} is described by the vertex
\begin{eqnarray}
V_{3A}=\frac{g}{2}\int \langle \partial_\nu
A_\mu,[A_\mu,A_\nu]\rangle d^4x\nonumber\\
={g}\int \partial_\nu A_\mu^a A_\mu^b A_\nu^c f^{abc} d^4x.
\end{eqnarray}

 \textbf{The gluon-ghosts interaction} is described by
\begin{eqnarray}
V_{A\bar{c}c}=-\frac{g}{2}\int \langle \partial_\mu \bar{c},[A_\mu,c]\rangle dx\nonumber\\
=-g \int \partial_\mu \bar{c}^a A_\mu^b c^c f^{abc} dx.
\end{eqnarray}
\indent \textbf{The fermion-gluon interaction}
\begin{eqnarray}
V_{A\bar{\psi}\psi}=ig\bar{\psi}\gamma_\mu
A_\mu\psi=-gA_\mu^a\bar{\psi}\gamma_\mu\Gamma^a\psi .
\end{eqnarray}
 Let us introduce the following notation for the Fourier
 transformation \(\tilde{f}(k)\):
\begin{eqnarray}
f(x)=\int e^{ikx}\tilde{f}(k)dk.
\end{eqnarray}
We have the following expression for the free gauge propagator
\begin{eqnarray}
\langle A_\mu^a(x) A_\nu^b(y)\rangle_0=\delta^{ab}\delta^{\mu\nu}
\frac{1}{(2\pi)^4} \int \frac{e^{ik(x-y)}}{k^2}dk,
\end{eqnarray}
for the free ghost propagator
\begin{eqnarray}
\langle\bar{c}^a(x) c^b(y)\rangle_0=\delta^{ab} \frac{1}{(2\pi)^4}
\int \frac{e^{ik(x-y)}}{k^2}dk,
\end{eqnarray}
and for the free fermion propagator
\begin{eqnarray}
\langle\bar{\psi}(x)\psi(x)\rangle_0=\delta^{ab}
\frac{1}{(2\pi)^4} \int \frac{e^{ik(x-y)}}{-\gamma_\mu k_\mu+m}dk.
\end{eqnarray}
In Fourier representation we have
\begin{eqnarray}
\langle \tilde{A}_\mu^a(k) \tilde{A}_\nu^b(k')\rangle_0
=\delta^{ab}\delta^{\mu\nu}\frac{1}{(2\pi)^4}\delta(k+k')\frac{1}{k^2},
\end{eqnarray}
\begin{eqnarray}
\langle\tilde{\bar{c}}^a(k) \tilde{c}^b(k')\rangle_0=\delta^{ab}
\frac{1}{(2\pi)^4} \delta(k+k') \frac{1}{k^2}.
\end{eqnarray}
Our aim is to define the gauge transformation on the Hopf algebra
of Feynman graphs. First of all we must prove the Slavnov ---
Tailor identities for individual diagrams.
\section{The Slavnov --- Taylor identities for individual diagrams}

\textbf{Definition.} Let \(v\) be a vertex of the diagram
\(\Gamma\). Suppose that \(n\) gluon lines come into \(v\), \(m'\)
fermion lines come from \(v\), \(m\) fermion lines comes into
\(v\), \(k\) ghost lines comes into \(v\) and \(k'\) ghost lines
comes from \(v\). Let
\begin{eqnarray}
w(x_1,...,x_n|y_1,...,y_m|z_1,...,z_{m'}|v_1,...,v_k|w_1,...,w_{k'}).
\end{eqnarray}
be an element of \(S_v\) (vertex operator) in coordinate
representation. We assign to each such operator the following
expression (Vick monomial)
\begin{eqnarray}
V=\int w(x_1,...,x_n|y_1,...,y_m|z_1,...,z_{m'}|v_1,...,v_k|w_1,...,w_{k'})\nonumber\\
\times
A(x_1)...A(x_n)\bar{\psi}(y_1)...\bar{\psi}(y_m)\psi(z_1)....\psi(z_{m'})\nonumber\\
\times\bar{c}(v_1)...\bar{c}(v_k)c(w_1)...c(w_{k'})\prod
\limits_{i=1}^{n}dx_i\prod \limits_{i=1}^{m}dy_i\prod
\limits_{i=1}^{m'}dz_i\prod \limits_{i=1}^{k}dv_i\prod
\limits_{i=1}^{k'}dw_i,
\end{eqnarray}
\(V\) is also called the vertex operator.

Let \(\omega=1+\alpha\) be an infinitezimal gauge transformation,
where  \(\alpha\) is an \(\mathfrak{g}\)-valued distribution such
that its Fourier transform is a finite linear combination of
\(\delta\)-functions
\begin{eqnarray}
\tilde{\alpha}(k)=\sum \limits_{i=1}^n c_i \delta(k-k_i).
\end{eqnarray}
The gauge variation \(\delta_\alpha V\) of \(V\) by definition is
a new vertex operator:
\begin{eqnarray}
\delta_\alpha V=  g \sum \limits_{i=1}^n
\int w(x_1,...,x_n|y_1,...,y_m|z_1,...,z_{m'}|v_1,...,v_k|w_1,...,w_{k'})\nonumber\\
\times
A(x_1)...[\alpha(x_i),A(x_1)]...A(x_n)\bar{\psi}(y_1)...\bar{\psi}(y_m)\psi(z_1)....\psi(z_{m'})\nonumber\\
\times\bar{c}(v_1)...\bar{c}(v_k)c(w_1)...c(w_{k'})\prod
\limits_{i=1}^{n}dx_i\prod \limits_{i=1}^{m}dy_i\prod
\limits_{i=1}^{m'}dz_i\prod \limits_{i=1}^{k}dv_i\prod
\limits_{i=1}^{k'}dw_i\nonumber\\
-g \sum \limits_{i=1}^m \int
w(x_1,...,x_n|y_1,...,y_m|z_1,...,z_{m'}|v_1,...,v_k|w_1,...,w_{k'})\nonumber\\
\times
A(x_1)...A(x_n)\bar{\psi}(y_1)...\bar{\psi}(y_i)\alpha(y_i)...\bar{\psi}(y_m)\psi(z_1)....\psi(z_{m'})\nonumber\\
\times\bar{c}(v_1)...\bar{c}(v_k)c(w_1)...c(w_{k'})\prod
\limits_{i=1}^{n}dx_i\prod \limits_{i=1}^{m}dy_i\prod
\limits_{i=1}^{m'}dz_i\prod \limits_{i=1}^{k}dv_i\prod
\limits_{i=1}^{k'}dw_i\nonumber\\
+ g \sum \limits_{i=1}^{m'} \int
w(x_1,...,x_n|y_1,...,y_m|z_1,...,z_{m'}|v_1,...,v_k|w_1,...,w_{k'})\nonumber\\
\times
A(x_1)...A(x_n)\bar{\psi}(y_1)...\bar{\psi}(y_m)\psi(z_1)...\alpha(z_i)\psi(z_i)...\psi(z_{m'})\nonumber\\
\times\bar{c}(v_1)...\bar{c}(v_k)c(w_1)...c(w_{k'})\prod
\limits_{i=1}^{n}dx_i\prod \limits_{i=1}^{m}dy_i\prod
\limits_{i=1}^{m'}dz_i\prod \limits_{i=1}^{k}dv_i\prod
\limits_{i=1}^{k'}dw_i\nonumber\\
+ g \sum \limits_{i=1}^{k'} \int
w(x_1,...,x_n|y_1,...,y_m|z_1,...,z_{m'}|v_1,...,v_k|w_1,...,w_{k'})\nonumber\\
\times
A(x_1)...A(x_n)\bar{\psi}(y_1)...\bar{\psi}(y_m)\psi(z_1)...\psi(z_{m'})\nonumber\\
\times\bar{c}(v_1)...\bar{c}(v_k)c(w_1)...[\alpha(w_i),c(w_i)]...c(w_{k'})\nonumber\\
\times \prod \limits_{i=1}^{n}dx_i\prod \limits_{i=1}^{m}dy_i\prod
\limits_{i=1}^{m'}dz_i\prod \limits_{i=1}^{k}dv_i\prod
\limits_{i=1}^{k'}dw_i.
\end{eqnarray}

It is easy to see that \(\delta_\alpha V \in S_v \).

 \textbf{Example 1.} The gauge variation of four-gluon vertex is
equal to zero.

 \textbf{Example 2.}
The gauge variation of the three gluon vertex is equal to
\begin{eqnarray}
\delta_\alpha
V_{3A}=\frac{g^2}{2}\langle[\partial_\nu\alpha,A_\mu],[A_\mu,A_\nu]\rangle.
\end{eqnarray}

 \textbf{Example 3.} The gauge variation of the vertex describing
 the gluon-fermion interaction is equal to zero.\newline
\indent \textbf{Example 4.} The gauge variation of vertex,
describing the ghost-fermion interaction is equal to
\begin{eqnarray}
\delta_\alpha
V_{\bar{c}cA}=-\frac{g}{2}\langle\partial_\mu\bar{c},[\alpha,[A_\mu,c]]\rangle.
\end{eqnarray}

Now we must define the \textbf{\(\xi\)}-insertion into the
vertecies and propagators.

 \textbf{\(\xi\)-insertion into the four-gluon vertex} is equal to
 zero.

\textbf{\(\xi\)-insertion into the three-gluon vertex} is equal to
\begin{eqnarray}
-\delta_\omega
V_{3A}=-\frac{g^2}{2}\langle[\partial_\nu\alpha,A_\mu],[A_\mu,A_\nu]\rangle.
\end{eqnarray}
\indent \textbf{Remark.} \(\xi\)-insertion into three-gluon vertex
is a minus gauge transformation of this vertex.\newline
 \indent \textbf{\(\xi\)-insertion into the ghost-gluon
 vertex} is equal to zero.\newline
 \indent \textbf{\(\xi\)-insertion into the fermion-gluon
vertex} is equal to zero.\newline
 \indent \textbf{\(\xi\)-insertion into the gluon line.} To obtain a \(\xi\)-insertion
 into the gluon line we must insert into this line the following two-photon vertex.
\begin{eqnarray}
\frac{g}{2}\langle\partial_\nu\partial_\mu\alpha,
[A_\mu,A_\nu]\rangle+\frac{g}{2}\langle\partial_\nu
A_\mu[\partial_\mu\alpha,A_\nu]\rangle +\frac{g}{2}\langle
\partial_\nu A_\mu,
[A_\mu,\partial_\nu\alpha]\rangle\nonumber\\
=-\frac{g}{2}\langle\Box
A_\mu[A_\mu,\alpha]\rangle-\frac{g}{2}\langle\partial_\mu
A_\mu,\partial_\mu[ \alpha,A_\mu]\rangle.
\end{eqnarray}
\indent \textbf{\(\xi\)-insertion into the ghost line.} To obtain
a \(\xi\)-insertion into the ghost line one must insert into this
line the following two-ghost vertex.
\begin{eqnarray}
-\frac{g}{2}\langle\partial_\mu\bar{c},[\partial_\mu\alpha,c]\rangle.
\end{eqnarray}
\indent \textbf{\(\xi\)-insertion into the fermion line.} To
obtain a \(\xi\)-insertion into the fermion line one must insert
the following two-fermion vertex into this line.
\begin{eqnarray}
ig\bar{\psi}(x)\gamma_\mu (\partial_\mu \alpha) \psi(x).
\end{eqnarray}

\indent\textbf{\(\eta\)-insertions.} We will see below that the
\(\eta\)-insertions comes from gauge variations of the action. Let
\(\omega=1+\alpha\) be an infinitesimal gauge transformation. The
gauge transformation of the action is equal to
\begin{eqnarray}
\delta_\alpha S=\delta_\alpha S_{G.F.}+\delta_\alpha
S_{F.P.},\nonumber\\
\delta_\alpha S_{G.F.}=-\frac{1}{2}\langle\partial_\mu
A_\mu,\Box\alpha\rangle+\frac{g}{2}\langle\partial_\mu
A_\mu,\partial_\mu [A_\mu,\alpha]\rangle, \nonumber\\
\delta_\alpha
S_{F.P.}=-\frac{1}{2}\langle\partial_\mu\bar{c},[\alpha,[\nabla_\mu,c]]\rangle.
\end{eqnarray}
\indent \textbf{\(\eta\) - insertion into the gluon line.} To
obtain a \(\eta\) - insertion into the gluon line we must to
insert the following two-gluon vertex into this line

\begin{eqnarray} \frac{g}{2}\langle\partial_\mu
A_\mu,\partial_\mu[\alpha,A_\mu]\rangle.
\end{eqnarray}

\textbf{Remark.} Note that the sun of \(\xi\) and \(\eta\)
insertions into  into the gluon line is equal to
\begin{eqnarray}
 \frac{g}{2}\langle\Box A_\mu,[\alpha,A_\mu]\rangle.
 \end{eqnarray}

 \textbf{\(\eta\) - insertion into
the ghost line.} To obtain a \(\eta\)-insertion into the ghost
line one must to insert the following two-ghost vertex into this
line
\begin{eqnarray}
-\frac{g}{2}\langle\partial_\mu\bar{c},[\alpha,\partial_\mu
c]\rangle.
\end{eqnarray}
\indent \textbf{Remark.} One can easily see that the sum of
\(\xi\)- and \(\eta\)-insertion into the ghost line is equal to
\begin{eqnarray}
\frac{g}{2}\langle \square \bar{c},[\alpha,c]\rangle.
\end{eqnarray}
\indent \textbf{\(\eta\) - insertion into the fermion-gluon
vertex} is equal to zero.

 \indent \textbf{\(\eta\) - insertion into the ghost-gluon
 vertex} replace the vertex operator \(-\frac{g}{2}\langle\partial_\mu\bar{c},[A_\mu,c]\rangle\)
 by
\begin{eqnarray}
\frac{g^2}{2}\langle\partial_\mu\bar{c},[\alpha,[A_\mu,c]]\rangle.
\end{eqnarray}
\indent \textbf{Remark.} Note that the \(\eta\)- insertion into
this vertex is equal to minus its gauge variation.

  \textbf{The Feynman rule for generating functional} \(\Gamma
 [A,\bar{\psi},\psi]\).

 To obtain the contribution from all one particle irreducible \(n\)-vertex diagrams into
 \(\Gamma[A,\bar{\psi},\psi]\) one must draw \(n\) points, then one must to replace
 each of this point by one of the vertecies from previous list, then we must connect this points by lines.
 We get diagrams. Then we must to each line assign a propagator etc.
 It is necessary to note that we do not identify topologically
 equivalent diagrams. The formalization of this procedure is
 simple and omitted.

 \textbf{Theorem 4.} \textsl{The Slavnov --- Taylor identity for
 individual diagrams. Let \(G\) be a one particle irreducible
 diagram without external ghost lines. Let \(G_\xi\) and \(G_\eta\) be
 diagrams, obtained from \(G\) by doing \(\xi\)- and \(\eta\)-
 insertion into some line or vertex of the diagram \(G\). Denote
 by \(\Gamma_G[A,\bar{\psi},\psi]\) the contribution into the
 generating functional, corresponding by \(G\). We have}

 \begin{eqnarray}
 \sum \limits_{\xi}\Gamma_{G_\xi} [A,\bar{\psi},\psi]+\sum \limits_{\eta}\Gamma_{G_\eta}
 [A,\bar{\psi},\psi]+\delta_\omega\Gamma_G[\omega A\omega^{-1},\bar{\psi}\omega^{-1},\omega
 \psi]=0. \label{ISTI}
 \end{eqnarray}
\textsl{ Here the first sum is over all \(\xi\)-insertions into
the
 diagram, and the second sum is over all \(\eta\)-insertions into \(\Gamma\).
}

\textbf{Proof.} Let us consider the sum of \(\xi\)- and \(\eta\)-
insertion into gluon line. We have shown that this sum is equal to
\(-g\square A_\mu^a[\alpha,A_\mu]^a\). Not that the free
propagator \(\langle A_\mu^a(x)A_\nu^a(x)\rangle_0\) is a
fundamental solution of the Laplace equation
\begin{eqnarray}
\square_x\langle A_\mu^a(x)
A_\nu^a(y)\rangle_0=-\delta(x-y)\delta^{a,b}\delta_{\mu,\nu}.
\end{eqnarray}
We see that \(\xi\)- and \(\eta\)- insertions into the gluon lines
leads to the gauge transformation of photon shoots which are the
ends of the line.

 Similarly one can see that \(\xi\)- and
 \(\eta\)-insertions into the ghost line leads to the gauge
 transformation of shoots which are the ends of the line.

The term
\begin{eqnarray}
\delta_\omega\Gamma_G[{}^\omega A,{}^\omega \bar{\psi},{}^\omega
\psi] \indent
\end{eqnarray}
leads to the gauge transformation of all shoots corresponding to
all external lines.

Let us now consider the gluon --- fermion vertex. We have seen
that all \(\xi\)- and \(\eta\)-insertions into the lines leads to
the gauge variation of all shoots of this vertex, i.e. to the
gauge variation of this vertex. The \(\xi\)- and
\(\eta\)-insertions into this vertex are equal to zero. But this
vertex is a gauge invariant. Therefore the sum of all gauge
variations of all shoots of this vertex and \(\xi\)- and
\(\eta\)-insertions into the vertex is equal to zero.

Now let us consider the three-gluon vertex. We have seen that all
the \(\xi\)- and \(\eta\)-insertions into the lines leads to the
gauge variation of all shoots of this vertex, i.e. leads to the
gauge transformation of this vertex. But \(\eta\)-insertion into
this vertex is equal to zero and \(\xi\)-insertion into this
vertex is equal to minus gauge variation of this vertex. Therefore
the sum of all gauge variations of all shoots of this vertex and
\(\xi\)- and \(\eta\)- insertions into this vertex is equal to
zero.

Similarly we can consider the gluon --- ghost and four-gluon
vertices.

Theorem is proved.

By definition the sum of all \(\xi\)- and \(\eta\)-insertions into
the fixed vertex \(v\) is called the \(\zeta\)-insertion into
\(v\). We have seen that \(\zeta\)-insertion into each vertex
coming from the action \(S=S_{YM}+S_F+S_{FP}+S_{GF}\) is precisely
a minus gauge variation of this vertex. We have proved the Slavnov
--- Taylor identity only for the diagrams coming from the action
\(S\). To define the gauge transformation on the algebra of
diagrams we must consider the diagrams containing arbitrary
vertexes. Therefore we define a \(\zeta\)-insertion into an
arbitrary vertex \(v\) as a minus gauge transformation of this
vertex. The following theorem holds.

\textbf{Theorem 5. (Generalized Slavnov --- Taylor equality.)}
\textsl{ For each one particle irreducible diagram \(G\) (with
arbitrary vertices) the following identity holds}:
\begin{eqnarray}
\delta_\omega\Gamma_G[\omega A\omega^{-1},
\bar{\psi}\omega^{-1},\omega \psi]+
 \sum \limits_{\zeta}\Gamma_{G_\zeta} [A,\bar{\psi},\psi]=0.
 \end{eqnarray}

 \textbf{Proof.} The proof of this theorem is a copy of the prove
 of previous theorem.

 \indent Now let us show how to derive the Slavnov --- Taylor identity from
 the Slavnov --- Taylor identity for the individual diagrams.
 To simplicity we consider only the case of pure Yang --- Mills
 theory.

 Let us summarize the identities (\ref{ISTI}) over all one-particle irreducible diagrams.
The sum over all \(\eta\) insertion is precisely a
\(-\delta_\omega\Gamma_\omega[A]\). The sum over all diagrams of
\(\delta_\omega \Gamma[\omega A \omega^{-1}]_G\) is equal to
\(\delta_\omega \Gamma[\omega A \omega^{-1}]\). Let us show that
the sum over all \(\xi\)-insertions of \(\Gamma[ A]_{G_\xi}\) is
equal to \( \int \frac{\delta\Gamma[A]}{\delta
A_\mu}\partial_\mu\alpha d^4 x \). If we prove this fact the
statement will be proved because
\begin{eqnarray}
 \delta_\omega \Gamma[{}^\omega A]=\int \frac{\delta
 \Gamma[A]}{\delta A_\mu}(\partial_\mu
 \alpha-[A,\alpha]_\mu)\nonumber\\
 =\delta_\omega\Gamma[\omega A \omega^{-1}]+\int \frac{\delta
 \Gamma[A]}{\delta A_\mu}{\partial_\mu \alpha}.
 \end{eqnarray}

We have the following representation for the generating functional
 \begin{eqnarray}
 \Gamma[A]=\sum \limits_n \frac{1}{n!} \sum \limits_m \frac{1}{m!}
 \int \Gamma^m_n(x_1,...,x_m)A(x_1)...A(x_m) dx_1...dx_m.
\end{eqnarray}
Here \(\Gamma_n^m\) is a sum of Feynman amplitude over all one
particle irreducible diagrams with \(n\) vertices and \(m\)
external lines (shoots). We suppose that the vertices and the
external lines are not identical. Let us represent \(\Gamma_n^m\)
as \(\Gamma_n^m=\sum \limits_{G_n} \Gamma^m_{G_n}\). Here the last
sum is over all one particle irreducible diagrams with \(n\)
vertices and  \(m\) external lines. We have
\begin{eqnarray}
\int \frac{\delta\Gamma[A]}{\delta A_\mu}\partial_\mu
\alpha\nonumber\\
 =\sum \limits_n \frac{1}{n!} \sum \limits_m
\frac{1}{(m-1)!}
 \int \sum
 _{G_n}\Gamma^m_{G_n}(x_1,...,x_m)\partial\alpha(x_1)...A(x_m).
 dx_1...dx_m,
 \end{eqnarray}
 We can rewrite the last formula as follows
\begin{eqnarray}
\int \frac{\delta\Gamma[A]}{\delta A_\mu}\partial_\mu
\alpha\nonumber\\
 =\sum \limits_n \frac{1}{(n)!} \sum \limits_m
\frac{1}{(m)!}
 \int \sum _{G_n}\Gamma^m_{G_n}(x_0,...,x_m)\partial\alpha(x_0)...A(x_m)
 dx_0...dx_m. \label{120}
 \end{eqnarray}
 Here we begin numerate vertices and external lines from zero.

 Let \(G_n^m\) be a one particle irreducible diagram with \(n\)
 vertices and \(m\) external lines. Let \(\xi\) be a
 \(\xi\)-insertion into some vertex or line. To each pair
 \((G_n^m,\xi)\) assign a diagram \((G_n^m)_\xi\) by doing a
 \(\xi\)-insertion. One can easily show what we can rewrite the
 right hand side of (\ref{120}) as follows
\begin{eqnarray}
\sum \limits_n \frac{1}{(n)!} \sum \limits_m \frac{1}{(m)!}
 \int \sum \limits_{(G_n^m,\xi)} \Gamma^{m}_{(G_{n}^m)_\xi}
 (x_1,...,x_m) A(x_1)...A(x_{m})
 dx_0...dx_m. \label{ST}
 \end{eqnarray}

 The right hand side of (\ref{ST}) is equal to
\(\sum \limits_G \sum \limits_\xi \Gamma_{G_\xi}[A]\). Therefore
the Slavnov
--- Taylor identity is proved.
\section{Gauge transformation on the Hopf algebra of diagrams}
At first we must give some definition. Let \(\Gamma\) be an one
particle irreducible diagram. Suppose that n gluon lines come into
\(\Gamma\), \(m\) fermion lines comes into \(\Gamma\) and \(m'\)
fermion lines come from \(\Gamma\), \(k\) ghost lines comes into
\(\Gamma\) and \(k'\) ghost lines comes from \(\Gamma\). Let
\begin{eqnarray}
f(x_1,...,x_n|y_1,...,y_m|z_1,...,z_{m'}|v_1,...,v_k|w_1,...,w_{k'}).
\end{eqnarray}
be an element of \(S_\Gamma\) in coordinate representation. We
assign to this element the following expression (vertex operator)
\begin{eqnarray}
V_f=\int w(x_1,...,x_n|y_1,...,y_m|z_1,...,z_{m'}|v_1,...,v_k|w_1,...,w_{k'})\nonumber\\
\times
A(x_1)...A(x_n)\bar{\psi}(y_1)...\bar{\psi}(y_m)\psi(z_1)....\psi(z_{m'})\nonumber\\
\times\bar{c}(v_1)...\bar{c}(v_k)c(w_1)...c(w_{k'})\prod
\limits_{i=1}^{n}dx_i\prod \limits_{i=1}^{m}dy_i\prod
\limits_{i=1}^{m'}dz_i\prod \limits_{i=1}^{k}dv_i\prod
\limits_{i=1}^{k'}dw_i.
\end{eqnarray}
Let \(\alpha\) be a \(\mathfrak{g}\)-valued distribution on
\(\mathbb{R}^4\) which Fourier transform has the form
 \begin{eqnarray}
\tilde{\alpha(k)}=c\delta(k-k_0)
\end{eqnarray}
By definition the gauge variation of \(V_f\) is a new vertex
operator
\begin{eqnarray}
\delta_\alpha V_f=g \sum \limits_{i=1}^n
\int w(x_1,...,x_n|y_1,...,y_m|z_1,...,z_{m'}|v_1,...,v_k|w_1,...,w_{k'})\nonumber\\
\times
A(x_1)...[\alpha(x_i),A(x_1)]...A(x_n)\bar{\psi}(y_1)...\bar{\psi}(y_m)\psi(z_1)....\psi(z_{m'})\nonumber\\
\times\bar{c}(v_1)...\bar{c}(v_k)c(w_1)...c(w_{k'})\prod
\limits_{i=1}^{n}dx_i\prod \limits_{i=1}^{m}dy_i\prod
\limits_{i=1}^{m'}dz_i\prod \limits_{i=1}^{k}dv_i\prod
\limits_{i=1}^{k'}dw_i\nonumber\\
 -g \sum \limits_{i=1}^m \int
w(x_1,...,x_n|y_1,...,y_m|z_1,...,z_{m'}|v_1,...,v_k|w_1,...,w_{k'})\nonumber\\
\times
A(x_1)...A(x_n)\bar{\psi}(y_1)...\bar{\psi}(y_i)\alpha(y_i)...\bar{\psi}(y_m)\psi(z_1)....\psi(z_{m'})\nonumber\\
\times\bar{c}(v_1)...\bar{c}(v_k)c(w_1)...c(w_{k'})\prod
\limits_{i=1}^{n}dx_i\prod \limits_{i=1}^{m}dy_i\prod
\limits_{i=1}^{m'}dz_i\prod \limits_{i=1}^{k}dv_i\prod
\limits_{i=1}^{k'}dw_i\nonumber\\
+ g \sum \limits_{i=1}^{m'} \int
w(x_1,...,x_n|y_1,...,y_m|z_1,...,z_{m'}|v_1,...,v_k|w_1,...,w_{k'})\nonumber\\
\times
A(x_1)...A(x_n)\bar{\psi}(y_1)...\bar{\psi}(y_m)\psi(z_1)...\alpha(z_i)\psi(z_i)...\psi(z_{m'})\nonumber\\
\times\bar{c}(v_1)...\bar{c}(v_k)c(w_1)...c(w_{k'})\prod
\limits_{i=1}^{n}dx_i\prod \limits_{i=1}^{m}dy_i\prod
\limits_{i=1}^{m'}dz_i\prod \limits_{i=1}^{k}dv_i\prod
\limits_{i=1}^{k'}dw_i\nonumber\\
+ g \sum \limits_{i=1}^{k'} \int
w(x_1,...,x_n|y_1,...,y_m|z_1,...,z_{m'}|v_1,...,v_k|w_1,...,w_{k'})\nonumber\\
\times
A(x_1)...A(x_n)\bar{\psi}(y_1)...\bar{\psi}(y_m)\psi(z_1)...\psi(z_{m'})\nonumber\\
\times\bar{c}(v_1)...\bar{c}(v_k)c(w_1)...[\alpha(w_i),c(w_i)]...c(w_{k'})\nonumber\\
\times\prod \limits_{i=1}^{n}dx_i\prod \limits_{i=1}^{m}dy_i\prod
\limits_{i=1}^{m'}dz_i\prod \limits_{i=1}^{k}dv_i\prod
\limits_{i=1}^{k'}dw_i.
\end{eqnarray}

It is easy to see that this definition is well defined i.e.
\(\delta_\alpha V_f=V_{{\delta_\alpha}f}\) for some unique
function \(\delta_\alpha f \in S_\Gamma\). Let \(\sigma \in
S^{'}_\Gamma\). By definition let \(\delta_\alpha(\sigma)\) be an
element of \(\sigma \in S^{'}_\Gamma\) such that
\begin{eqnarray}
\langle \delta_\alpha(\sigma),f\rangle=\langle\sigma,
\delta_\alpha(f)\rangle.
\end{eqnarray}
Here \(\langle \sigma,f\rangle\) means the value of functional
\(\sigma\) on \(g\).

\textbf{Definition.} Let \(\alpha\) be a \(\mathfrak{g}\)-valued
distribution on \(\mathbb{R}^4\) such that its Fourier transform
has the form
\begin{eqnarray}
\tilde{\alpha}(k)=\sum \limits_{j=1}^n c_j\delta(p_j-p).
\end{eqnarray}
By definition the gauge transformation \(\delta_\alpha\) on
\(\mathcal{H}\) is its derivative as an algebra defined on
generators as follows
\begin{eqnarray}
\delta_\alpha((\Gamma,\sigma))=\delta'_\alpha((\Gamma,\sigma))+\delta''_\alpha((\Gamma,\sigma)),
\label{I1}
\end{eqnarray}
where we put
\begin{eqnarray}
\delta'_\alpha((\Gamma,\sigma))=\sum_{\zeta \in
\Gamma}(\Gamma_\zeta,\sigma), \label{I2}
\end{eqnarray}
and
\begin{eqnarray}
\delta''_\alpha((\Gamma,\sigma))=(\Gamma,\delta_\alpha(\sigma))
\label{I3}.
\end{eqnarray}
\textbf{Theorem 6.} \textsl{The gauge transformation is a
derivative of \(\mathcal{H}\), i.e.}
 \begin{eqnarray}
 \Delta\circ\delta_\alpha=(\mathbf{1}\otimes\delta_\alpha+\delta_\alpha\otimes\mathbf{1})\circ\Delta,
 \nonumber\\
\varepsilon \circ\delta_\alpha=0,\nonumber\\
S\circ\delta_\alpha=\delta_\alpha\circ S.\nonumber
\end{eqnarray} \textbf{Proof.} We have
\begin{eqnarray}
\Delta\circ\delta_\alpha((\Gamma,\sigma))\nonumber\\
=\Delta\circ\delta'_\alpha((\Gamma,\sigma))+\Delta\circ\delta''_\alpha((\Gamma,\sigma)).
\end{eqnarray}
it is evidence that
\begin{eqnarray}
\Delta\circ\delta''_\alpha((\Gamma,\sigma))=(\mathbf{1}\otimes\delta''_\alpha)\circ\Delta((\Gamma,\sigma)).
\end{eqnarray}
Therefore we must calculate:
\begin{eqnarray}
\Delta\circ\delta'_\alpha((\Gamma,\sigma))=\Delta(\sum
\limits_\zeta(\Gamma_\zeta,\sigma)).
\end{eqnarray}
We have
 \begin{eqnarray}
 \sum \limits_\zeta \Delta((\Gamma_\zeta,\sigma))\nonumber\\
 =(\Gamma,\sigma)\otimes\mathbf{1}
 +\mathbf{1}\otimes(\Gamma,\sigma)
+ \sum \limits_\zeta\sum
 \limits_{\gamma_\alpha\subset\Gamma_{\zeta}} \gamma_\alpha
 \otimes({\Gamma_\zeta}/{\gamma_\alpha},\sigma). \label{Th}
\end{eqnarray}
But the last sum is equal to
\begin{eqnarray}
\sum \limits_\zeta\sum
 \limits_{\gamma_\alpha\subset\Gamma_{\zeta}}  \gamma_\alpha
 \otimes({\Gamma_\zeta}/{\gamma_\alpha},\sigma)\nonumber\\
=\sum \limits_{{\gamma}_\alpha\subset\Gamma}\sum \limits_{\zeta
\in \gamma} ({\gamma}_\alpha)_\zeta
\otimes({\Gamma}/{{\gamma}_\alpha},\sigma) \nonumber\\
+\sum \limits_{{\gamma}_\alpha\subset\Gamma}\sum
\limits_{\zeta}^{'} {\gamma}_\alpha
 \otimes({\Gamma_\zeta}/{{\gamma}_\alpha},\sigma).\label{th1}
\end{eqnarray}
Here \('\) at the last sum means that all the \(\zeta\)-insertions
into the sum are the \(\zeta\)-insertions into the vertices or
lines of \(\Gamma\) which do not belong to \(\gamma\).

Let us transform the first term in the right hand side of
(\ref{th1}). We have
\begin{eqnarray}
 \sum
\limits_{{\gamma}_\alpha\subset\Gamma}\sum \limits_{\zeta}
({\gamma}_\alpha)_\zeta
\otimes({\Gamma}/{{\gamma}_\alpha},\sigma)\nonumber\\
 =\sum \limits_{{\gamma}_\alpha\subset\Gamma}\sum \limits_{\zeta}
({\gamma}_\alpha)_\zeta
\otimes({\Gamma}/{{\gamma}_\alpha},\sigma)\nonumber\\
+\sum \limits_{{\gamma}_\alpha\subset\Gamma}
\delta_\alpha''({\gamma}_\alpha)
\otimes({\Gamma}/{{\gamma}_\alpha},\sigma)
\nonumber\\
-\sum \limits_{{\gamma}_\alpha\subset\Gamma}
\delta_\alpha''({\gamma}_\alpha)
\otimes({\Gamma}/{{\gamma}_\alpha},\sigma). \label{AB}
\end{eqnarray}
By definition of \(\delta_\alpha\) the sum of first and second
terms in the right hand side of (\ref{AB}) is equal to
\begin{eqnarray}
(\delta_\alpha\otimes\mathbf{1})\sum
\limits_{{\gamma}_\alpha\subset\Gamma}{\gamma}_\alpha
 \otimes({\Gamma}/{{\gamma}_\alpha},\sigma).
\end{eqnarray}
The last term in right hand side is equal to
\begin{eqnarray}
 \sum \limits_{\gamma_\alpha\subset\Gamma} \sum \limits_\zeta^{''}
\gamma_\alpha\otimes((\Gamma/\gamma_\alpha)_\zeta,\sigma).
\end{eqnarray}
Here \(''\) means that all \(\zeta\)-insertion are made into the
vertices of \(\Gamma/\gamma_\alpha\) obtained by replacing of all
connected components of \(\gamma\) by vertices. As result we have

 \begin{eqnarray}
 \Delta\circ\delta_\alpha((\Gamma,\sigma))=((\delta_\alpha\otimes\mathbf{1})+
 (\mathbf{1}\otimes\delta_\alpha))\circ\Delta((\Gamma,\sigma)).
\end{eqnarray}
It follows from this fact that
\begin{eqnarray}
\Delta\circ\delta_\alpha=(\mathbf{1}\otimes\delta_\alpha+\delta_\alpha\otimes\mathbf{1})\circ\Delta.
\end{eqnarray}
 Similarly one can prove that
 \begin{eqnarray}
\varepsilon \circ\delta_\alpha=0,\nonumber\\
S\circ\delta_\alpha=\delta_\alpha\circ S.\nonumber
\end{eqnarray}
The theorem is proved.

\indent \textbf{Remark.} Below we will consider only characters
\(U\) such that \(U((\Gamma,l^{\alpha'}_{\Gamma}))\neq0\), \(
 l^{\alpha'}_{\Gamma} \in B_\Gamma^{'}\) only for finite number of
 elements \(\alpha \in B_\Gamma^{'}\). For any two such characters \(U_1\) and \(U_2\) its product
 \(U_1\star U_2\) well defined.

  \textbf{Remark.} Let \(\mathcal{G}\) be a linear space of all
\(\mathfrak {g}\)-valued functions on \(\mathbb{R}^4\) of the form
\begin{eqnarray}
\sum \limits_{i=1}^N a_ie^{ik_ix} \label{form}.
\end{eqnarray}
\(\mathcal{G}\) is a Lie algebra with respect to the following Lie
brackets
\begin{eqnarray}
[\alpha_1,\alpha_2](x)=[\alpha_1(x),\alpha_2(x)].
\end{eqnarray}

 \textbf{Theorem 7.} \textsl{Gauge transformation \(\delta\) is a
 homomorphism from \(\mathcal{G}\) to the  Lie algebra of all
 derivatives of \(\mathcal{H}\)}.

 \textbf{Remark.} We can define a gauge transformation
 \(\delta_\alpha\) on comodule \(M\) by using the formulas similar
 to (\ref{I1}, \ref{I2}, \ref{I3}). We find that \(\delta_\alpha\) is a derivative of comodule
 \(M\) i.e.

 \begin{eqnarray}
 \Delta\circ
 \delta_\alpha(x)=(\mathbf{1}\otimes\delta_\alpha+\delta_\alpha\otimes\mathbf{1})\circ\Delta(x).
 \end{eqnarray}
\indent \textbf{Definition.} Let \(\alpha\) be a
\(\mathfrak{g}\)-valued function on \(\mathbb{R}^4\) of the form
(\ref{form}). We say that character \(U\) is gauge invariant if
\begin{eqnarray}
\delta^\star_\alpha(U):= U\circ\delta_\alpha=0
\end{eqnarray}
\indent \textbf{Remark.} Let \(M'\) is an algebraically dual
module of \(M\) over the group algebra of \(G\). Dimensionally
regularized Feynman amplitude define an element \(m \in M'\). We
say that \(m \in M'\) is gauge invariant if
\(m\circ\delta_\alpha=0\) \(\forall \alpha\) of the form
(\ref{form}).

\textbf{Theorem 8.} \textsl{The element \(m \in M'\) corresponding
to dimensionally regularized Feynman amplitude is gauge
invariant.}

\textbf{Proof.} This theorem follows from the Slavnov --- Taylor
identities for diagrams.

 \textbf{Theorem 9.} \textsl{The set of all gauge invariant characters of \(G\) is a group}.

 \textbf{Proof.} Let \(U_1\) and \(U_2\) be gauge
invariant characters. We have:
\begin{eqnarray}
U_1\star U_2\circ\delta_\alpha=U_1\otimes
U_2\circ\Delta\circ\delta_\alpha\nonumber\\
=(U_1\otimes
U_2)\circ((\mathbf{1}\otimes\delta_\alpha)+(\delta_\alpha\otimes\mathbf{1}))\circ\Delta\nonumber\\
=(U_1\circ\delta_\alpha) \star U_2)+U_1\star
(U_2\circ\delta_\alpha)=0.
\end{eqnarray}
So the product of two gauge invariant character is a gauge
invariant character. Let us prove that for each character \(U\)
its inverse character \(U^{-1}\) is a gauge invariant. Indeed
\begin{eqnarray}
U^{-1}\circ\delta_\alpha=U\circ S\circ\delta_\alpha=
U\circ\delta_\alpha\circ S=0.
\end{eqnarray}\newline
Theorem is proved.

\textbf{Definition.} Character \(U\) is called gauge invariant up
to degree \(n\) if \(\delta^{\ast}(U)((\Gamma,\sigma))=0\) for all
diagrams \(\Gamma\) which contain at most \(n\) vertices.

\textbf{Remark.} Let \(C\) be gauge invariant character up degree
\(n-1\) and \(U\) be a character. One can prove that
 \begin{eqnarray}
  \{ U (\bullet )
 +  \sum \limits_{\emptyset
 \subset \gamma_{\alpha} \subset \bullet } C( \gamma_{\alpha})
 U( \bullet / {\gamma_{\alpha}}) \}\delta_{\alpha}((\Gamma,\sigma))\nonumber \\
 =\{\delta_{\alpha}^{\ast} U (\bullet )+ \sum \limits_{\emptyset
 \subset \gamma_{\alpha} \subset \bullet } C( \gamma_{\alpha})
 ( \delta_{\alpha}^{\ast}U) (\bullet /{\gamma_{\alpha}})\}((\Gamma,\sigma)).
 \end{eqnarray}
 For any diagram \(\Gamma\) which contain at most \(n\) vertices.

 \textbf{Definition.} Let \(D\) be a open set in
 \(\mathbb{C}\). The function \(U_z\), \(z\rightarrow U_z\) is
 called continuous, holomorphic, etc. if

a) \(\forall X \in \mathcal{H}\) \(U_z(X))\) is a continuous,
holomorphic, ect. in \(D\).

b) For any diagram \(\Gamma\), \(l_\Gamma^{\alpha'} \in B_\Gamma\)
\(
 U_z((\Gamma,\alpha'))\equiv 0)\) in \(D\) for all elements \(l_\Gamma^{\alpha'} \in B_\Gamma\)
 except some finite subset.

 \textbf{Riemann --- Hilbert problem.} Let \(U_{z}\) --- be a character holomorphic,
 in some small punctured neigbourhood of zero \(O \setminus \{0\}\).
 It is aimed to find two characters \(R_{z}\) and \(C_{z}\), holomorphoc in \(z\) in \(O\)
 and \(\mathbf{C}\setminus \{0\}\) respectively such that the following identities holds

 \begin{equation}
 R_{z}=C_{z} \star U_{z} \label{raz}
 \end{equation}
in \(O\setminus \{0\} \) and \(C_{z} \rightarrow \mathbf{1}\) if \(z
\rightarrow \infty\). The pair \((R_z, C_z)\) is called the Birkhoff
decomposition of \(U_z\).

The uniquiness of solution follows from the Liuville theorem.

\textbf{Theorem 10 (Connes --- Kreimer).} \textsl{The Riemann ---
Hilbert problem for group of characters has a solution.}

 \textbf{Proof.} One can find the following explicit formulas for
 the solution of the problem:
 \begin{equation} C_{z}((\Gamma,\sigma))=-\mathbf{T}(U_{z}((\Gamma,\sigma))+ \sum
 \limits_{\emptyset \subset \gamma_{\alpha} \subset \Gamma}
 C_{z}((\gamma_{\alpha},\sigma)) U_{z}((\Gamma / \gamma_{\alpha},\sigma))),
 \end{equation}
 \begin{equation}
 R_{z}((\Gamma,\sigma))=(1-\mathbf{T})(U_{z}((\Gamma,\sigma))+ \sum
 \limits_{\emptyset \subset \gamma_{\alpha} \subset \Gamma}
 C_{z}(\gamma_{\alpha}) U_{z}((\Gamma / \gamma_{\alpha},\sigma))).
 \end{equation}
 By definition an operator \(\mathbf{T}\) assigns to each Laurent
 series
\begin{eqnarray}
\sum \limits_{j=-n}^{\infty} a_i z^i
\end{eqnarray}
the following polynomial on \(z^{-1}\)
\begin{eqnarray}
\sum \limits_{j=-n}^{-1} a_i z^i.
\end{eqnarray}

 \textbf{The Riemann --- Hilbert problem on \
\(M'\).} Let \(m_z \in  M'\) be an element of \(M'\) holomorphic
in some punctured neigbourhood of zero \(\mathcal{O}\setminus
\{0\}\). This means that \(\forall (\Gamma,\sigma) \in M\)
\(m_z((\Gamma,\sigma))\) holomorphic in \(\mathcal{O}\setminus
\{0\}\).  It is aimed to find element \(C_z \in G\) and \(m^+_z
\in M'\) holomorphic in \( \overline{\mathbb{C}}\setminus\{0\}\)
and \(\mathcal{O}\) respectively such that in
\(\mathcal{O}\setminus \{0\}\) the following identities hold
\begin{eqnarray}
m_z^+=C_z\star m_z \label{dva},
\end{eqnarray}
and
\begin{eqnarray}
C_z(\infty)=\varepsilon.
\end{eqnarray}

\textbf{Remark.} If \(m_z\) corresponds to dimensionally regularized
Feynman amplitudes then the existence of solution follows from the
Bogoliubov --- Parasiuk theorem.

  \textbf{Theorem 11.} \textsl{If the solution of the Riemann --- Hilbert problems
  (\ref{raz}, \ref{dva})
  exist and the data of these problems are gauge invariant then the
  elements of their Birkhoff decompositions \((R_z, C_z)\), (\((m_z^+,C_z)\)) are gauge invariant
  too.}

 \textbf{Proof.} The proof follows from the fact that \(\mathbf{T}\) commutes with
 \(\delta_\alpha\) and from remark to theorem 1.

 \textbf{Remark.} Let \(m_z\) be an element of \(M'\) corresponding to the set of dimensionally
 regularized Feynman amplitudes. Character \({m}_z^+\) corresponds to renormalized Feynman amplitudes
 and \(C_z\) corresponds to counterterms.
 \section{Independence of vacuum expectation value of gauge invariant functional of
  the chose of gauge condition}

  Let us show (on physical level of curiosity) that expectation value of gauge
  invariant functional does not depend
  of the chose of gauge condition.

  Let us denote non-renormalized Green function as
\begin{eqnarray}
\langle ...A...\bar{c}...c...\rangle=\int DAD\bar{c}Dc
e^{-S}[A,\bar{c},{c}]...A...\bar{c}...c...
\end{eqnarray}
For simplicity we consider the case of pure Yang --- Mills theory.

Denote renormalized Green functions as
\begin{eqnarray}
\langle...A...\bar{c}...c...\rangle_R=\int DAD\bar{c}Dc
\{e^{-S}[A,\bar{c},{c}]\}_R...A...\bar{c}...c...
\end{eqnarray}

Let \(F_R[J]\) be a generating functional for renormalized
connected Green function, and \(\Gamma_R[J]\) be its Legendre
transform. One can prove that \(\Gamma_R[A]\) is a generating
functional for one particle irreducible renormalized Green
functions.

Let \(\omega=1+\alpha\) be an infinitezimal gauge transformation.
Now let us use gauge condition \(g[{}^\omega A]=0\) instead the
gauge condition \(g[A]=0\). Let us denote the expectation value
corresponding to the new gauge condition \(g[{}^\omega A]=0\) by
\(\langle\rangle'\). We find that
\begin{eqnarray}
\langle...A...\bar{c}...c\rangle'_R=\int DAD\bar{c}Dc
\{e^{-S}[{}^\omega A,{}^\omega
\bar{c},{}^\omega{c}]\}_R...A...\bar{c}...c. \label{153}
\end{eqnarray}
Recall that under the gauge transformation the ghosts transforms
as follows
\begin{eqnarray}
c\mapsto\omega c\omega^{-1},\nonumber\\
\bar{c}\mapsto \bar{c}.
\end{eqnarray}
In (\ref{153}) we must at first make the gauge transformation and
 then make the renormalization. The Legendre transformation assign
 to the connected generating functional for green functions
 (\ref{153}) the functional \((\Gamma_R)_\omega[A]\). It follows from
 the gauge invariance of renormalized Feynman amplitudes that
\begin{eqnarray}
(\Gamma_R)_\omega[A]=\Gamma_R[{}^\omega A].
\end{eqnarray}
But inverse Legendre transformation assigns to
\(\Gamma_R[{}^\omega A]\) the following set of Green functions
\begin{eqnarray}
\langle...A...\bar{c}...c...\rangle''_R=\int DAD\bar{c}Dc
\{e^{-S}_R[{}^\omega A,{}^\omega
\bar{c},{}^\omega{c}]\}...A...\bar{c}...c....
\end{eqnarray}
Here one must at first make renormalization of \(e^{-S}\) and then
make the gauge transformation.

Let \(F[A,\bar{c},{c}]\) be an enough regular gauge invariant
functional. Consider its expectation value for the new gauge
condition  \(g[{}^\omega A]=0\). We have
\begin{eqnarray}
\langle F[A,\bar{c},{c}]\rangle'=\int DAD\bar{c}Dc
\{e^{-S}[{}^\omega A,{}^\omega \bar{c},{}^\omega{c}]\}_R
F[A,\bar{c},{c}]\nonumber\\
=\int DA^{\omega^{-1}}D\bar{c}Dc^{\omega^{-1}}
\{e^{-S}[A,\bar{c},{c}]\}_R
F[{}^{\omega^{-1}}\bar{c},{}^{\omega^{-1}}{c}]\nonumber\\
=\int DA D \bar{c} Dc \{ e^{-S} \}_R [A,\bar{c},{c}]
F[A,\bar{c},{c}]=\langle F[A,\bar{c},{c}]\rangle.\nonumber\\
\end{eqnarray}
Therefore
\begin{eqnarray}
\langle F[A,\bar{c},{c}]\rangle '=\langle F[A,\bar{c},{c}]\rangle.
\end{eqnarray}
The statement is proved.
 Note that we have proved the following statement.

 \textbf{Proposition.} For any enough regular gauge invariant functional \(F[A]\) the following identity holds:
\begin{eqnarray}
 \int DADDc  \bar{c}\{\int dx^4 \alpha(x)\frac{\delta  e^{-S}[{}^\omega A ,{}^\omega \bar{c},{}^\omega c]}
 {\delta\alpha(x)}\}_R
F[A,\bar{c},{c}],\nonumber\\
\omega=1+\alpha.
\end{eqnarray}

Now let \(g'[A]=0\) be a new gauge condition such that the
difference \(g'[A]-g[A]\) is infinitely small. Let \(A\) be a
field configuration such that \(g[A']=0\). Let \(A'\) be a field
configuration such that \(g'[A']=0\). Suppose that \(A\) and
\(A'\) belong to the same class of gauge equivalent field
configuration. There exists an infinitely small function
\(\alpha[A](x)\) of \(x\) which is gauge invariant functional of
\(A\) such that
\begin{eqnarray}
A'=A+\nabla_A\alpha.
\end{eqnarray}
This statement follows from
\begin{eqnarray}
\rm det \mit\left \|\frac{\delta g[{}^\omega
A]}{\delta\omega}\right \|\neq 0.
\end{eqnarray}
There exist enough many functions in this class of the form
 \begin{eqnarray}
\alpha[A](x)=\sum \limits_{i=1}^{n} f_i(x) G_i[A],
\end{eqnarray}
where \(f_i(x)\) are well \(\mathfrak{g}\)-valued functions on
\(\mathbb{R}^4\) and \(G_i[A]\) are enough regular functionals of
\(A\). Let us find the variation \(\delta\langle F[A]\rangle\)
corresponding to the variation \(\delta g[A]=g'[A]-g[A]\) of gauge
function. We have

\begin{eqnarray}
\delta\langle F[A]\rangle\nonumber\\
\int DAD \bar{c}Dc\{\int d^4 x\frac{\delta e^{-S}[{}^\omega A
,{}^\omega\bar{c},{}^\omega {c}]}{\delta\alpha(x)}\sum
\limits_{i=1}^{n}
f_i(x) G_i[A]\}_R F[A,\bar{c},{c}]\nonumber\\
=\sum  \limits_{i=1}^{n} \int DAD \bar{c}Dc\{\delta_{f_i}
e^{-S}[{}^\omega A ,{}^\omega\bar{c},{}^\omega {c}] G_i[A]\}_R
F[A,{\bar{c}},{c}].
\end{eqnarray}
\(G_i[A]\) are enough regular functionals, so
\begin{eqnarray}
\{...G_i[A]\}_R=\{...\}_R G_i[A].
\end{eqnarray}
Therefore we have
\begin{eqnarray}
\delta\langle F[A]\rangle\nonumber\\
 =\sum  \limits_{i=1}^{n}
\int DAD \bar{c}Dc\{\delta_{f_i} e^{iS}[{}^\omega A
,{}^\omega\bar{c},{}^\omega {c}]\}_R G_i[A]F[A,\bar{c},{c}]=0.
\end{eqnarray}
The statement is proved.

To prove the fact  that \(S\)-matrix is unitary it is enough to
prove that \(S\)-matrix is gauge independent. And to prove gauge
independence of \(S\) one can use similar arguments.
\section{Conclusion}
In this work we have given the generalization of the Connes ---
Kreimer method in renormalization theory to the case of nonabelian
gauge theories. We have introduced the Hopf algebra of diagrams
which generalize the corresponding construction of Connes and
Kreimer. We have defined a gauge transformation on this Hopf
algebra.

We have obtained three main results. The first one is that the
gauge transformation is a derivation of the Hopf algebra of
diagrams. The second one is that the set of all gauge invariant
characters is a group. The third one is that the Riemann ---
Hilbert problem has a gauge invariant solution if the data of this
problem is gauge invariant. We have shown how to simply prove that
renormalized \(S\)-matrix is gauge invariant.

I wold like to thank I.V. Volovich for the problem setup and A.V.
Zayakin for very useful discussions.

\end{document}